\documentclass[12pt]{article}
\usepackage{graphicx}

\begin{document}

\baselineskip=14pt plus 1pt minus 1pt

\begin{center}

{\large \bf Exactly separable version of the Bohr Hamiltonian with
the Davidson potential}

\bigskip\bigskip

{Dennis Bonatsos$^{a}$\footnote{e-mail: bonat@inp.demokritos.gr},
E. A. McCutchan$^{b}$\footnote{e-mail:
elizabeth.ricard-mccutchan@yale.edu}, N.
Minkov$^{c}$\footnote{e-mail: nminkov@inrne.bas.bg}, R.F.
Casten$^{b}$\footnote{e-mail: richard.casten@yale.edu}, P.
Yotov$^{c}$\footnote{e-mail: pyotov@inrne.bas.bg}, D.
Lenis$^{a}$\footnote{e-mail: lenis@inp.demokritos.gr}, D.
Petrellis$^{a}$\footnote{e-mail: petrellis@inp.demokritos.gr}, I.
Yigitoglu$^{d}$\footnote{e-mail: yigitoglu@istanbul.edu.tr} }
\bigskip

{$^{a}$ Institute of Nuclear Physics, N.C.S.R. ``Demokritos'',}

{GR-15310 Aghia Paraskevi, Attiki, Greece}

{$^{b}$ Wright Nuclear Structure Laboratory, Yale University,}

{New Haven, Connecticut 06520-8124, USA}

{$^{c}$ Institute of Nuclear Research and Nuclear Energy,}

{72 Tzarigrad Road, 1784 Sofia, Bulgaria}

{$^{d}$ Hasan Ali Yucel Faculty of Education, Istanbul
University,}

{TR-34470 Beyazit, Istanbul, Turkey}

\end{center}

\bigskip\bigskip
\centerline{\bf Abstract} \medskip

An exactly separable version of the Bohr Hamiltonian is developed
using a potential of the form $u(\beta)+u(\gamma)/\beta^2$, with
the Davidson potential $u$($\beta$) = $\beta^2+\beta_0^4/\beta^2$
(where $\beta_0$ is the position of the minimum) and a stiff
harmonic oscillator for $u$($\gamma$) centered at
$\gamma=0^{\circ}$. In the resulting solution, called exactly
separable Davidson (ES-D), the ground state band, $\gamma$ band
and $0_2^+$ band are all treated on an equal footing. The
bandheads, energy spacings within bands, and a number of interband
and intraband $B(E2)$ transition rates are well reproduced for
almost all well-deformed rare earth and actinide nuclei using two
parameters ($\beta_0$, $\gamma$ stiffness). Insights regarding the
recently found correlation between $\gamma$ stiffness and the
$\gamma$-bandhead energy, as well as the long standing problem of
producing a level scheme with Interacting Boson Approximation
SU(3) degeneracies from the Bohr Hamiltonian, are also obtained.


Section: Nuclear structure

\newpage

\section{Introduction} 

The Bohr Hamiltonian \cite{Bohr} has been at the foundation of the
collective model description of nuclei for over fifty years.
Numerous solutions have been proposed since its derivation by
choosing different forms of the potential $V$($\beta$,$\gamma$)
and solving the corresponding eigenvalue equation either
analytically or approximately.  Recently, this approach has
undergone renewed interest, due in part, to the development of the
concept of critical point symmetries (CPS).  These models, E(5)
\cite{IacE5} and X(5) \cite{IacX5}, are special solutions of the
Bohr Hamiltonian designed to describe nuclei at the critical point
of the shape/phase transition between vibrational and
$\gamma$-soft or axially symmetric deformed structures,
respectively.

In E(5) \cite{IacE5}, a $\gamma$-independent potential of the form
$u(\beta)$ is used, leading to exact separation of $\beta$ from
$\gamma$ and the Euler angles \cite{Wilets}, while in X(5)
\cite{IacX5} a potential of the form $u(\beta)+u(\gamma)$ is
assumed, leading to an approximate separation of variables in the
special case of $\gamma \approx 0^{\circ}$, achieved by
$u(\gamma)$ being a stiff harmonic oscillator centered at
$\gamma=0^{\circ}$. In both E(5) and X(5) an infinite square well
potential is used as $u(\beta)$, in accordance with growing
evidence from microscopic calculations\cite{Meng,Sheng,Fossion}
that the potential at the transition point between different
shapes should be flat. Model predictions for energy spectra and
$B(E2)$ transition rates are parameter free (up to overall scale
factors) in E(5) \cite{IacE5}, while in X(5) the predictions
related to the ground state band and the excited $0^+$ bands are
parameter free, but $\gamma$ bands contain the stiffness parameter
of the $\gamma$ oscillator \cite{IacX5,Bijker}.

It often happens that a successful, but simple, model or approach
spawns new generations of related approaches.  This is especially
the case if, despite its success, the data reveal certain, albeit
perhaps small, discrepancies with the simple approach.  A classic
case of this is the simple formula for rotational spectra
\cite{Bohr}, which led to a myriad of alternate formulas (see for
example Refs. \cite{harris,buck,ejiri,lipas}), usually more and
more parameters and, not surprisingly, working better. Of course,
each such case ultimately entails a judgement as to whether the
additional complications are worth the improved descriptions they
yield.

The case of critical point symmetries is no exception.  Despite
their simplicity (square wells in $\beta$ along with flat or
harmonic oscillator potentials in the $\gamma$ degree of freedom)
and their success in describing transitional nuclei, it was
immediately recognized that there were important discrepancies
with the data as well. One, for example, occurs in X(5) where the
predicted energy spacings in the excited $0^+_2$ band are far too
large \cite{CZX5,Kruecken}.

Since the advent of these CPS, a number of alternate geometrical
models have been proposed and their predictions worked out.  Some
of these share with X(5) an extreme economy of parameters, others
have one additional parameter.  These models can all be solved
exactly, either analytically or numerically. Some are, in fact,
essentially identical to the CPS but are solved exactly, while
others involve alternate, presumably more realistic potentials.
One example, which we shall refer to occasionally, is the
so-called Confined Beta Soft (CBS) model \cite{CBS1,CBS2} which
takes as its starting point from X(5) but allows the inner wall to
move out to the radius of the outer wall. As the inner wall moves,
the spectra change smoothly from X(5) to a pure rotor. Other
potentials \cite{Fort,Fort2,Heyde}, which we shall not consider,
utilize triaxial shapes with non-zero values for the minimum of
the potential in $\gamma$.

It is the purpose of this paper to explore a few of the most
promising geometrical models, to compare their predictions with
each other and with the data.  We deal only with the axially
symmetric case at present, that is, nuclei whose potentials in
$\gamma$ are of harmonic oscillator type with a minimum at
$\gamma$ = $0^{\circ}$.

These models can be grouped into three classes: one, called
X$_{\textrm {ex}}$(5) is simply an exact numerical solution
\cite{Caprio} of the Hamiltonian of X(5), without the approximate
separation of $\beta$ and $\gamma$ variables used in Ref.
\cite{IacX5}. This type of exact solution has now become
tractable, using the novel techniques introduced in Refs.
\cite{RoweI,RoweII,RoweIII}. It is worth mentioning that in
X$_{\textrm {ex}}$(5), the $\gamma$ stiffness parameter is
involved in all bands, while in X(5) the ground and $\beta$ bands
are independent of the $\gamma$ stiffness parameter .

The other two classes each take advantage of a kind of potential
that is exactly separable from the start.  Such potentials have
the form \cite{Wilets}

\begin{equation}
u_{ES}(\beta, \gamma) = u(\beta) + \frac{u(\gamma)}{\beta^2}
\end{equation}

\noindent where ES stands for exactly separable.  The first of
these uses the same $u(\beta)$ and $u(\gamma)$ as X(5) itself -
that is, the square well in $\beta$ and a $\gamma$ dependent
potential given by a harmonic oscillator in $\gamma$.  This is the
so-called ES-X(5) solution \cite{ESX5}.

The second group of the exactly separable class of potentials uses
the Davidson potential \cite{Dav} in $\beta$, namely

\begin{equation}
u(\beta) = \beta^2 + \frac{\beta_0^4}{\beta^2}
\end{equation}

\noindent where $\beta_0$ is the free parameter and gives the
position of the minimum of the potential in $\beta$. The use of
the Davidson potential with an approximate separation of variables
has been discussed in Refs. \cite{varPLB,varPRC}. In the present
work, we examine the Davidson potential with an exactly separable
potential, which we call exactly separable Davidson (ES-D). By
including a harmonic oscillator potential in $\gamma$, analytic
solutions can be derived in this form to describe well-deformed,
axially symmetric nuclei. These Davidson potentials, along with
the X(5) potential, are illustrated in Fig. 1(a) for both the
approximate separation of variables (left) and the exactly
separable cases (right). In Fig. 1(b), the Davidson potential in
just the $\beta$ degree of freedom is illustrated for a few values
of the $\beta_0$ parameter.

Before proceeding to a detailed discussion of the present solution
of the Bohr Hamiltonian, it is useful to put the present work in a
context of other solutions to the Bohr Hamiltonian. This
Hamiltonian has been solved analytically in the $\gamma$-unstable
case [$u(\beta,\gamma)=u(\beta)$] using the Davidson potential of
Eq. (2) as the $\beta$-potential \cite{Elliott}, showing, that
with increasing values of the $\beta_0$ parameter, a transition
from the spherical vibrator to a rigid non-spherical
$\gamma$-unstable structure occurs. The link provided by O(5)
between the $\gamma$-unstable geometrical model and the O(6) limit
of the Interacting Boson Approximation (IBA) model \cite{IA} has
also been previously studied \cite{Evans}. Later, it was shown
\cite{Bahri} that the above mentioned $\gamma$-unstable Bohr
Hamiltonian with the Davidson potential is characterized by the
symmetry SU(1,1)$\times$SO(5), with SO(5) due to rotational
invariance in the five-dimensional collective space, and with
SU(1,1) due to the Davidson potential. If the potential is allowed
to also depend on $\gamma$, no algebraic solution has been found,
but it has been shown that numerical calculations converge much
more rapidly in an SO(5) basis with $\beta_0\neq 0$ than in the
usual spherical basis with $\beta_0=0$
\cite{RoweI,RoweII,RoweIII}: the relevant SO(5) spherical
harmonics having been calculated in Ref. \cite{RoweII}. The
correspondence between this approach, called the algebraic
collective model \cite{RoweIII}, and the different limiting
symmetries of the Interacting Boson Approximation (IBA) model
\cite{IA} has been studied in Refs. \cite{Thiamova1,Thiamova2}.
This powerful method has been recently extended \cite{RoweIV} to
the SU(1,1)$\times$SO(N) case.

In view of the above, the present work is an analytic, special
solution of the Bohr Hamiltonian with a Davidson $\beta$-potential
appropriate for axially symmetric prolate deformed nuclei (since
the $\gamma$-potential is taken to possess a steep minimum at
$\gamma=0$), while the earlier solutions of Refs.
\cite{Elliott,Bahri} refer to a Davidson $\beta$-potential in a
$\gamma$-unstable framework. As a result, the present solution
will turn out to be appropriate for the description of
well-deformed axially symmetric nuclei, which comprise the bulk of
well-deformed nuclei, while the solution of Refs.
\cite{Elliott,Bahri} is appropriate for those $\gamma$-unstable
nuclei between spherical and moderately deformed cases.

There are several advantages in the present ES-D solution which we
will consider in detail. As mentioned above, no approximation is
involved in the separation of variables. As a result, all bands
(ground, $\gamma$, and $\beta$) are treated on an equal footing
depending on two parameters, the Davidson parameter $\beta_0$
(which is the location of the minimum of the potential) and the
stiffness $c$ of the $\gamma$ oscillator. Finally, the $\beta^2$
term in the potential solves the spacing problem in the $\beta$
band that plagues the infinite square well solutions. Of course,
with a minimum in $\gamma$ at 0$^{\circ}$ and a relatively steep
potential in $\gamma$, the model is applicable only to axially
deformed rotational nuclei.

Despite this constraint, it will be shown that the present
solution provides good results for the spectra and $B(E2)$
transition rates of almost all well deformed rare earth and
actinide nuclei. Furthermore, it provides insights regarding the
recently found correlation \cite{Hinke} between $\gamma$ stiffness
and the $\gamma$-bandhead energy, as well as the long standing
problem of producing an Interacting Boson Approximation (IBA)
SU(3) degenerate level scheme \cite{IA} within the framework of
the Bohr Hamiltonian.

\section{The ES-D model} 

Our starting point is the original Bohr Hamiltonian \cite{Bohr}
\begin{equation}\label{eq:e1}
H = -{\hbar^2 \over 2B} \left[ {1\over \beta^4} {\partial \over
\partial \beta} \beta^4 {\partial \over \partial \beta} + {1\over
\beta^2 \sin 3\gamma} {\partial \over \partial \gamma} \sin 3
\gamma {\partial \over
\partial \gamma} - {1\over 4 \beta^2} \sum_{k=1,2,3} {Q_k^2 \over \sin^2
\left(\gamma - {2\over 3} \pi k\right) } \right] +V(\beta,\gamma),
\end{equation}
where $\beta$ and $\gamma$ are the usual collective coordinates,
while $Q_k$ ($k=1$, 2, 3) are the components of angular momentum
in the intrinsic frame, and $B$ is the mass parameter.

We assume that the reduced potential, $u = 2B V /\hbar^2$, can be
separated into two terms of the form

\begin{equation}
u(\beta, \gamma) = u(\beta) + \frac{u(\gamma)}{\beta^2}
\end{equation}

\noindent as in Refs. \cite{Wilets,Fort,Fort2,Heyde} where the
Schr\"odinger equation can then be separated exactly into two
equations.

For the potential in $\gamma$ we use a harmonic oscillator

\begin{equation}
u(\gamma)= (3c)^2 \gamma^2
\end{equation}

\noindent and $u(\beta)$ is taken as the Davidson potential
\cite{Dav,Elliott,Bahri}

\begin{equation}\label{eq:e7}
u(\beta)=\beta^2 + {\beta_0^4 \over \beta^2},
\end{equation}

\noindent where $\beta_0$ denotes the position of the minimum of
the potential.  As described in Appendix I, the resulting energy
eigenvalues are given by

\begin{equation}\label{eq:e13}
E_{n,L} = 2n+1 + \sqrt{ { L(L+1)-K^2\over 3} +{9\over 4} +
\beta_0^4 + 3C (n_\gamma+1) }, \qquad n=0,1,2,\ldots
\end{equation}

For $K=0$ one has $L=0$, 2, 4, \dots, while for $K\neq 0$ one
obtains $L=K$, $K+1$, $K+2$, \dots

Bands occurring in this solution, characterized by ($n$,
$n_\gamma$), include the ground state band $(0,0)$, the
$\beta_1$-band $(1,0)$, the $\gamma_1$-band $(0,1)$, and the first
$K=4$ band $(0,2)$. The relative position of all levels depends on
the parameters $\beta_{\circ}$ and $C$. ($C=2c$ is used in order
to keep equations similar to those in Refs. \cite{IacX5} and
\cite{ESX5}.) All bands are treated on equal footing
\cite{IacCam}, in analogy with the SU(3) limit of the Interacting
Boson Model \cite{IA}.

Details on the calculation of $B$($E2$) transition strengths are
described in Appendix II. We note that the $u(\gamma)$ potential
used in the Bohr equation has to be periodic, because of
coordinate symmetry constraints \cite{Bohr}. In Ref. \cite{Caprio}
both the proper periodic potential $(1-\cos 3\gamma)$ and the
approximate form $\gamma^2$, appropriate for small $\gamma$, have
been used, yielding similar results. A more detailed study of this
issue has been recently carried out \cite{Raduta}, leading to the
use of spheroidal or Mathieu functions. Periodic
$\gamma$-potentials involving $\cos 3\gamma$ have been used in an
early solution involving a harmonic oscillator for the
$\beta$-potential \cite{Jean}, as well as more recently in the
framework of the algebraic collective model
\cite{RoweI,Thiamova1,Thiamova2}, where their treatment is
tractable because $\cos 3\gamma$ is, within a constant, an SO(5)
spherical harmonic with $v=3$ and $L=0$ (where $v$ the seniority
and $L$ the angular momentum) \cite{RoweI}. The potential $\csc^2
3 \gamma$, which is the partner of the infinite well potential in
supersymmetric quantum mechanics \cite{SUSYQM}, has also been used
recently \cite{DeBaerde} in the Bohr Hamiltonian for triaxial
nuclei. It is certainly of interest to examine the consequences of
the use of periodic $\gamma$ potentials in the present approach in
subsequent work.

In the present paper we are going to follow Ref. \cite{Bijker},
normalizing $\Delta K=0$ transitions to $2_1^+\to 0_1^+$, and
$\Delta K=2$ transitions to $2_\gamma^+ \to 0_1^+$. In this way
normalization difficulties vanish.

The spectrum and $B(E2)$ transition strengths of ES-X(5) are
described for completeness in Appendix III.

\section{Numerical results and comparison to experiment} 

\subsection{Energy ratios} 

In  Fig. 2(a), the $R_{4/2}=E(4_1^+)/E(2_1^+)$ ratio as a function
of the parameter $C$ is shown for the ES-D solution (for a few
values of the Davidson parameter $\beta_{\circ}$) and ES-X(5), as
well as for the exact numerical solution [X$_{\textrm {ex}}$(5)]
of Ref. \cite{Caprio}. The parameter $C$ is connected to the
parameter $a$ of the exact numerical solution \cite{Caprio}
through $C={2\over 3}\sqrt{a}$. In Figs. 2(b) and (c), the ratios
$R_{0/2}=E(0_{\beta}^+)/E(2_1^+)$ and
$R_{2/2}=E(2_{\gamma}^+)/E(2_1^+)$, corresponding to the
normalized $\beta$ and $\gamma$ bandhead energies, respectively,
are shown for the same solutions.  The ES-D solution with
$\beta_0=0$ corresponds to the ES-X(5)-$\beta^2$ solution of Ref.
\cite{ESX5}.

From Fig. 2(a), it is clear that the ES-D and ES-X(5) solutions
are appropriate mostly for well-deformed nuclei, while the exact
numerical solution \cite{Caprio} is also applicable to less
deformed nuclei (including the $a=200$ case [$C=9.428$] which
gives results similar to the original X(5) model \cite{IacX5}).
This difference is due to the $\beta^2$ term in the potential
$u(\beta)+u(\gamma)/\beta^2$ used in the exactly separable cases.
Within the ES-D solution, the rotational limit of $R_{4/2}=10/3$
is closely approached already for $\beta_0 = 4$.

As seen in Fig. 2(b), the normalized $\beta$ bandhead energy,
$R_{0/2}$, has a large dependence on the parameter $\beta_0$ and
shows less variation with the stiffness parameter $C$,
particularly for large $\beta_0$ values.  This dependence is
reversed for the normalized $\gamma$ bandhead energy, $R_{2/2}$,
which varies only slightly for different $\beta_0$ values, but has
a large dependence on the $C$ parameter. As a result, the
$R_{2/2}$ and $R_{0/2}$ lines cross in ES-D at values of $C$
increasing with $\beta_0$. These lines also cross in the numerical
solution \cite{Caprio}, but they do not cross in ES-X(5). This
point will be further discussed in the next subsection.

Concerning the results of the exact numerical solution X$_{\textrm
{ex}}$(5) of Ref. \cite{Caprio} used for comparisons in this and
in subsequent sections, our assignment of levels to a particular
band follows the same as given in Ref. \cite{Caprio}. In
particular, $2_\gamma^+$ corresponds to $2_2^+$ for $a=0-450$ and
to $2_3^+$ for $a=500-1000$, while $2_\beta^+$ corresponds to
$2_3^+$ for $a=0-450$ and to $2_2^+$ for $a=500-1000$. Similarly,
$4_\gamma^+$ corresponds to $4_2^+$ for $a=0-650$ and to $4_3^+$
for $a=700-1000$, while $4_\beta^+$ corresponds to $4_3^+$ for
$a=0-650$ and to $4_2^+$ for $a=700-1000$. These assignments are
related to avoided crossings, as explained in Ref. \cite{Caprio}.

\subsection{Relative spacings within different bands and relative positions
of bandheads} 

In Fig. 3(a) the energy ratio

\begin{equation}\label{eq:e41}
R_{2 \beta}={E(2_{\beta}^+)-E(0_{\beta}^+) \over E(2_1^+)}
\end{equation}

\noindent is shown for the solutions under discussion. The ratio
is exactly 1 in the case of ES-D, irrespective of the value of the
Davidson parameter $\beta_0$. This is due to the oscillator term
in the Davidson potential which gives equal rotational spacings in
the ground state band and the $\beta$ band.  Thus, the same holds
for the energy ratio
\begin{equation}\label{eq:e42}
R_{4 \beta}={E(4_{\beta}^+)-E(2_{\beta}^+)   \over
E(4_1^+)-E(2_1^+)},
\end{equation}

\noindent shown in Fig. 3(b). In Fig. 3(c) the energy ratio

\begin{equation}\label{eq:e43}
R_{\gamma}= {E(4_{\gamma}^+)-E(2_{\gamma}^+) \over
E(4_1^+)-E(2_1^+)}
\end{equation}

\noindent is shown, while in Fig. 3(d) the energy ratio

\begin{equation}\label{eq:e44}
 R_{\beta \gamma}=R_{0/2}-R_{2/2} =
{E(0_{\beta}^+)-E(2_{\gamma}^+) \over E(2_1^+)}
\end{equation}

\noindent is given. Abrupt changes in the predictions of
X$_{\textrm {ex}}$(5) are due to the avoided crossings of
($2_2^+$, $2_3^+$) and ($4_2^+$, $4_3^+$).

Experimental data for the energy ratios shown in Fig. 3 are
exhibited in Fig. 4.  Since the current solution is only
applicable to well-deformed nuclei, the data included in Fig. 4 is
limited to $A$ $>$ 100 and $R_{4/2}$ $>$ 3.00. Comparing Figs.
3(a) and 4(a) we see that in terms of the energy ratio $R_{2
\beta}$, which compares the level spacing within the $\beta$ band
to the level spacing within the ground state band, most nuclei
exhibit a ratio slightly less than 1.0.  This feature is most
closely reproduced by the ES-D solution which predicts a ratio of
exactly 1.0, independent of the value of the Davidson parameter
$\beta_0$. The predictions of the ES-X(5) solution for $R_{2
\beta}$ are higher by 50\% or more, and the predictions of
X$_{\textrm {ex}}$(5) are even higher. (This is the well known
problem of overprediction of the spacing of the $\beta$ band in
the X(5) model by a factor close to two \cite{CZX5,Kruecken},
which can be resolved by replacing the infinite well potential by
a potential with linear sloped walls \cite{sloped}.) The same is
seen for the energy ratio $R_{4 \beta}$ in Figs. 3(b) and 4(b).

From Fig. 4(c), it is clear that the majority of the data for the
energy ratio $R_{\gamma}$, which compares the level spacing within
the $\gamma$ band to the level spacing within the ground state
band, is centered around values of 1.0.  The predictions for
$R_{\gamma}$ from each of the solutions overlap and are consistent
with the range observed in the data.  The ES-D solution gives the
largest range of predictions, since this is a more flexible model
(2 parameters) compared with the single parameter X$_{\textrm
{ex}}$(5) and ES-X(5) solutions.  Overall, all three solutions
yield reasonable predictions for the $\gamma$-band spacings in
deformed nuclei.

The experimental energy ratio $R_{\beta \gamma}$, which is related
to the relative positioning of the $\beta$ and $\gamma$ bandhead
energies, exhibits a wide range of values spanning positive to
negative, as shown in Fig. 4(d).  As a result, we expect that the
solutions exhibiting both positive and negative values for this
ratio, namely the ES-D solution for not very high values of the
Davidson parameter $\beta_0$ and X$_{\textrm {ex}}$(5), should
better reproduce this feature.

Summarizing the above observations, all three solutions under
consideration, ES-D, ES-X(5) and X$_{\textrm {ex}}$(5), are found
to give reasonable predictions for the $\gamma$-band spacing,
while the ES-D solution yields predictions which most closely
reproduce the $\beta$-band spacing of most deformed nuclei.  The
ES-D solution (for not very high values of the parameter
$\beta_0$) and the X$_{\textrm {ex}}$(5) solution appear to
reproduce the relative positions of the $\beta$ and $\gamma$
bandhead energies in a number of nuclei. Thus, the ES-D solution
provides the flexibility to describe a wide range of observables
(spacings within the $\beta$ and $\gamma$ bands, relative position
of the $\beta$ and $\gamma$ bandheads) with not very large values
of the Davidson parameter $\beta_0$.

\subsection{$B(E2)$ ratios} 

Having examined the main features of the energy spectra, we turn
now to the study of the characteristics of the $B(E2)$ transition
rates. As mentioned in Sec. 2, in order to avoid normalization
problems, $\Delta K=0$ transitions will be normalized to the
$2_1^+\to 0_1^+$ transition, while $\Delta K=2$ transitions will
be normalized to the $2_\gamma^+ \to 0_1^+$ transition, as in Ref.
\cite{Bijker}.  We include in this comparison the predictions of
the original X(5) solution as well as the U(5) and SU(3) limits of
the IBA.  X(5) predictions for  ground $\to$ ground, $\beta \to
\beta$, and $\beta \to $ ground transitions are taken from Ref.
\cite{IacX5}, while X(5) predictions for $\gamma \to \gamma$,
$\gamma \to $ ground, and $\gamma \to \beta$ transitions are taken
from Ref. \cite{Bijker}. SU(3) predictions for the  ground$ \to $
ground and $\beta \to \beta$ transitions are obtained with the
standard quadrupole operator of the IBA \cite{IA}, while
predictions for $\beta \to $ ground and $\gamma \to $ ground
transitions are obtained with the extended quadrupole operator
containing the extra term $(d^\dagger \times \tilde s+ s^\dagger
\times \tilde d)^{(2)}$ \cite{IA}.

Intraband transitions within the ground state and $\beta$ band are
shown in Fig. 5.  Within the ground state band, Fig. 5(a), the
ES-D predictions lie in between X(5) and SU(3) for most values of
$C$. Again, for $\beta_0$ values of 4 and larger, the SU(3) limit
is almost exactly achieved. For the transitions within the $\beta$
band, shown in Figs. 5(b),(c), the ES-D predictions lie between
U(5) and SU(3), approaching the latter with increasing values of
$\beta_0$. In the case of the ratio $B$($E$2;
$2_{\beta}^+\rightarrow0_{\beta}^+$) / $B$($E$2;
$2_1^+\rightarrow0_1^+$), the X(5) predictions lie below the SU(3)
value, and not between the U(5) and SU(3) values, as might have
been expected, since they are related to the transition between
U(5) and SU(3). As with the energy spectra, the $\beta$ band
predictions again exhibit the largest differences between ES-D and
X(5).

Transitions from the $\gamma$ band are shown in Fig. 6. The ES-D
predictions are consistently close to X(5) or intermediate between
X(5) and the SU(3) limit.  This is particularly true for the
branching ratios from the $\gamma$ band given in Figs. 6(b),(c).

The $\beta$ band to ground band transitions are shown in Fig. 7.
For the transition from the $0_2^+$ state to the ground state
$2_1^+$, the predictions of ES-D are intermediate between X(5) and
SU(3) for most values of $C$.  The decay from the $2_{\beta}^+$
state to the ground state band, Figs. 7(b),(c), shows some
variation between the models, but all are similar in magnitude.
These small differences become more evident when branching ratios
are considered, as in Fig. 7(d).  For the ratio, $B$($E$2;
$2_{\beta}^+\rightarrow4_1^+$) / $B$($E$2;
$2_{\beta}^+\rightarrow0_1^+$), the X(5) predictions are nearly an
order of magnitude larger than the SU(3) ratio (also the Alaga
ratio) of 2.6.  The ES-D predictions are again intermediate
between X(5) and SU(3).

The $\gamma$ band to $\beta$ band transitions are shown in Fig. 8.
The predictions of ES-D for growing $\beta_0$ approach X(5).

In summary, in (almost) all cases the ES-D predictions lie in
general between the X(5) and SU(3) predictions, with SU(3) already
approached at $\beta_0 = 4$.

\subsection{Fits to specific nuclei} 

A search has been made to find nuclei for which the ground state,
$\beta$, and $\gamma$ bands (up to the point of backbending or
upbending in each band) can be well reproduced by the ES-D
solution. Since the ES-D solution is appropriate only for deformed
nuclei, the search was constrained to nuclei with $R_{4/2}$ $>$
3.00. Considering all such nuclei in the rare-earth and actinide
regions, we find that almost all nuclei with a known $0_2^+$ and
$2_{\gamma}^+$ state can be well described in terms of energies by
ES-D, as shown in Table 1. The quality measure
\begin{equation}\label{eq:e99}
\sigma = \sqrt{ { \sum_{i=1}^n (E_i(exp)-E_i(th))^2 \over
(n-1)E(2_1^+)} },
\end{equation}
used for evaluating the rms fits performed, remains less than one
in most cases.  Out of the $\sim$ 60 nuclei which meet the above
criteria, there are only two cases where the ES-D solution does
not provide a good description of all the bandheand energies,
namely $^{152}$Sm and $^{154}$Gd. These exceptions are not
surprising, since these nuclei are well described by the X(5)
model, which uses a ``flat-bottomed'' potential in the $\beta$
degree of freedom.  The ES-D solution, on the other hand,
incorporates a potential which is much ``stiffer'' in the $\beta$
degree of freedom. Thus, discrepancies between the data and the
ES-D solution are expected in transitional nuclei and, indeed, may
be used to point to nuclei with flat potentials in the $\beta$
degree of freedom.

Several $B(E2)$ ratios obtained with ES-D using the same
parameters as given in Table 1 are shown in Table 2, which
includes all nuclei of Table 1 for which nontrivial information on
relevant $B(E2)$s is experimentally known \cite{NDS}. More
detailed level schemes for $^{156}$Gd and $^{232}$Th are shown in
Fig. 9, as examples of the quality of the ES-D solution to
reproduce detailed spectra.

As seen in Table 2, the intraband $B(E2)$ ratios within the ground
state bands are reproduced quite well for a majority of the
nuclei, despite the fact that $B(E2)$ values have not been taken
into account in the fitting procedure.  Also, the theoretical
$\gamma \to $ ground $B(E2)$ ratios are in very good agreement
with the experiment values. However, the theoretical $\gamma \to $
ground $B(E2)$ strengths, when normalized to the $2_1^+\to 0_1^+$
transition, are much lower than the experimental ones. This could
be due to the normalization difficulties mentioned at the end of
Sec. 2, which disappear if ratios of $\gamma \to $ ground
transitions are used. Moreover, the theoretical interband $ \beta
\to$ ground $B(E2)$ values are consistently an order of magnitude
higher than the experimental values.

\subsection{Bandheads} 

The ability of the present model to reproduce the general
experimental trends of $R_{0/2} = E(0_{\beta}^+)/E(2_1^+)$ and of
$R_{2/2} = E(2_{\gamma}^+)/E(2_1^+)$ as a function of $R_{4/2} =
E(4_1^+)/E(2_1^+)$ is shown in Figs. 10(a),(b). Predictions of the
ES-X(5), Confined Beta Soft (CBS) \cite{CBS1,CBS2} and X$_{\textrm
{ex}}$(5) solutions are also shown for comparison. From Eq. (7) it
is clear that the energy levels of the ground state and $\beta$
bands depend only on the parameter combination $\beta_0^4 + 3C$,
thus in Fig. 10(a) only one curve appears for ES-D, with
$\beta_0^4 + 3C$ increasing from left to right. From the same
equation it is also clear that the levels of the $\gamma$ band
depend on the parameter combination $\beta_0^4+ 6C$. As a result,
different curves are obtained for ES-D in Fig. 10(b) by fixing
$\beta_0$ to different values and varying the $C$ parameter.

The predictions for $R_{0/2}$ as a function of $R_{4/2}$ are more
less the same for the ES-D, ES-X(5) and CBS solutions.  The
X$_{\textrm {ex}}$(5) predictions for $R_{0/2}$ are slightly
higher and above the overall trend of the data.

As discussed previously, in the CBS solution and other
X(5)-related solutions, the bandhead energy of the $\gamma$ band
depends on a free parameter.  In the present exactly separable
(ES) solutions, it is treated on an equal footing as the $\beta$
bandhead energy. The plot of $R_{2/2}$ vs. $R_{4/2}$ reveals that
a large set of data corresponds to the ES-D region with $\beta_0$
between 2 and 4. The same set is also described quite well by the
ES-X(5) curve. Figure 10(b) also reveals that the predictions of
the ES-D solution for the $\gamma$ bandhead energy are only in
agreement with the data for $R_{4/2}$ values larger than 3.0. This
is again related to the present solution being applicable only to
axially symmetric well-deformed nuclei, since when the parameter
$C$ becomes too small, the approximation of an axially symmetric
potential is no longer valid. On the other hand, X$_{\textrm
{ex}}$(5) provides a better description of $R_{2/2}$ for $R_{4/2}$
values between 2.6 and 3.0.

\subsection{Gamma-stiffness} 

In Ref. \cite{Hinke} a correlation has been found between the
gamma stiffness of the potential and the ratio $R_{2/2}=
E(2_{\gamma}^+)/E(2_1^+)$, with the gamma-stiffness increasing
stronger than linearly as a function of $R_{2/2}$. In the present
model, the gamma-stiffness coefficient $(3c)^2$ is shown as a
function of $R_{2/2}$ in Fig. 11(a). It is evident that a stronger
than linear increase is seen, which varies little with $\beta_0$,
at least for reasonable values of the latter, as indicated from
Table 1.  The specific points corresponding to the rare earth and
actinide nuclei of Table 1 are shown in Fig. 11(b), exhibiting the
same trend.

A short discussion is now in place on the qualitative
correspondence between the two parameters ($\beta_0$, $C$) of the
present solution and those of the usual two-parameter IBA-1
Hamiltonian \cite{Zamfir66,Werner}

\begin{equation}\label{eq:IBA}
H(\zeta,\chi) = C \left[ (1-\zeta) \hat n_d -{\zeta\over 4 N_B}
\hat Q^\chi \cdot \hat Q^\chi\right],
\end{equation}

\noindent where $\hat n_d = d^\dagger \cdot \tilde d$, $\hat
Q^\chi = (s^\dagger \tilde d + d^\dagger s) +\chi (d^\dagger
\tilde d)^{(2)},$ $N_B$ is the number of valence bosons, and $C$
is a scaling factor. The above Hamiltonian contains two
parameters, $\zeta$ and $\chi$, with the parameter $\zeta$ ranging
from 0 to 1, and the parameter $\chi$ ranging from 0 to
$-\sqrt{7}/2=-1.32$. The IBA dynamical symmetries are given by
$\zeta=0$, any $\chi$ for U(5), $\zeta=1$, $\chi=-\sqrt{7}/2$ for
SU(3), and $\zeta=1$, $\chi=0$ for O(6). As remarked in Ref.
\cite{Hinke}, stiffness is proportional to the IBA parameter
$\chi$. Thus, in the present case $(3c)^2$ roughly corresponds to
$|\chi|$. On the other hand, we have already seen that increasing
$\beta_0$ leads to the SU(3) limit, thus $\beta_0$ is in
qualitative correspondence to $\zeta$. It should be emphasized,
however, that while the IBA Hamiltonian of Eq. (13) can cover the
whole region from U(5) ($R_{4/2}=2$)  to SU(3) ($R_{4/2}=3.33$),
the ES-D solution provides reasonable results only in the narrow
region of $R_{4/2}$ between 3.0 and 3.33.

\subsection{Occurrence of SU(3) degeneracy} 

A long standing problem has been deriving from the Bohr
Hamiltonian a spectrum similar to that of the SU(3) limit of the
Interacting Boson Approximation (IBA) model \cite{IA}. The main
features of the spectrum should be:

a) The energy spacings among the $2^+$, $4^+$, $6^+$, \dots levels
within the ground, $\beta$ and $\gamma$ bands should be identical.

b) Furthermore, the $2^+$, $4^+$, $6^+$, \dots   levels of the
$\beta$ and $\gamma$ bands should be degenerate.

In the present model, the spacings within the ground and $\beta$
bands are identical, because of the oscillator term in the
$u(\beta)$ potential, as already seen in subsection 3.2. It is
therefore sufficient to examine the conditions under which the
$2^+$, $4^+$, $6^+$, \dots levels of the $\beta$ and $\gamma$
bands are degenerate.

From Eq. (7) it is trivial to see that the energy spacings in the
$\beta$ and $\gamma$ bands become equal for any $L$ if $C=4/9$
(since in this case the $3C$ term in the $\beta$ band is
counterbalanced by the $-K^2/3 +6C$ term in the $\gamma$ band,
which has $K=2$). However, this observation is of little physical
significance, since the values of $C$ appropriate for actual
nuclei, appearing in Table 1, are considerably higher.

Figs. 3(b), (c) indicate that in general the spacings within the
$\gamma$ band are lower than the spacings within the $\beta$ band
by about 20\% for most $\beta_0$ and $C$ values of interest. Thus
within the present solution one can only hope to reproduce a
situation with approximate degeneracy for the first few even
levels of the $\beta$ and $\gamma$ bands.

From Eq. (7), the requirement $E(2_{\beta}^+)= E(2_{\gamma}^+)$
leads to the condition $ 9 C^2 -80 C -16 \beta_0^4 -356/9=0$,
while the requirement $E(4_{\beta}^+)=E(4_{\gamma}^+)$ leads to
the condition $ 9 C^2 - 80 C - 16 \beta_0^4 -1028/9=0$. Similar
conditions occur from the requirement $E(L_{\beta}^+)=
E(L_{\gamma}^+)$ for higher $L$. These conditions can be
approximately satisfied simultaneously only for very large values
of $\beta_0$, which are outside the region of physical interest
according to the $\beta_0$ values appearing in Table 1, since too
high $\beta_0$ would result in too high a $0_{\beta}^+$ bandhead
energy.

In this way one is led to consider what happens for a fixed value
of $R_{0/2}=E(0_{\beta}^+)/E(2_1^+)$. In this case Eq. (7) easily
leads to

\begin{equation}\label{eq:e31}
 3C = \left( {R_{0/2}\over 2} - {1\over R_{0/2}}\right) ^2 -\beta_0^4
-{9\over 4}.
\end{equation}

\noindent Thus for a given $R_{0/2}$ one can minimize with respect
to $\beta_0$ the rms deviation between the even levels of the
$\beta$ and $\gamma$ bands

\begin{equation}\label{eq:e32}
\sigma_{\beta,\gamma}(L_{max}) = \sqrt{ {1\over L_{max}/2-1}
\sum_{L=2}^{L_{max}} \left( {E(L_{\beta}^+)-E(L_{\gamma}^+) \over
E(2_1^+)} \right)^2 },
\end{equation}

\noindent the value of $C$ obtained for each $\beta_0$ from Eq.
(15). Numerical results shown in Table 3 indicate that a
reasonable degree of degeneracy is obtained for $L_{max} =10$ and
$R_{0/2}\geq 15$, which is of physical interest, since the
$R_{0/2}$ values in Table 1 extend up to 27~. In Table 4, the
results of the fit to $^{232}$Th, corresponding to the values
reported in Table 1 are given. In the case of $^{232}$Th, which is
very close to the $R_{0/2}=15$ case reported in Table 3, one can
see that $\sigma_{\beta,\gamma}^{th}(L_{\max}=10)=1.142$, while
$\sigma_{\beta,\gamma}^{exp}(L_{\max}=10)=0.593$. Therefore,
although the overall fit is quite good, the degree of degeneracy
obtained from theory is less than the one indicated by experiment.
One could conclude that the present solution does contain
parameter pairs which correspond to an approximate degeneracy of
the low-lying even levels of the $\beta$ and $\gamma$ bands, while
at the same time the spacings within the $\beta$ band are
identical to the spacings within the ground band, however the
problem of reproducing a SU(3) spectrum from the Bohr Hamiltonian
remains conceptually open.

\subsection{Alhassid-Whelan arc of regularity} 

It has been recently suggested that an experimental confirmation
\cite{arcexp} of the Alhassid-Whelan arc of regularity \cite{AW},
connecting the U(5) and SU(3) symmetries in the symmetry triangle
\cite{Casten} of the Interacting Boson Approximation (IBA) model
\cite{IA} is manifested in nuclei in which the $\beta$ and
$\gamma$ bandheads, $0^+_{\beta}$ and $2^+_{\gamma}$, are nearly
degenerate. From Eq. (7) the requirement $E(0_{\beta}^+)=
E(2_{\gamma}^+)$ leads to the condition $ 9 C^2 -68 C -16
\beta_0^4 -224/9=0$. Given the fact that $C$ has to be
nonnegative, this condition leads to $C= (34+\sqrt{1380+144
\beta_0^4})/9$. Among the nuclei listed in Table 1, the ones
satisfying the condition $ | E(2_{\gamma}^+)-E(0_{\beta}^+)
|/E(2_{\gamma}^+) \leq 0.05 $ \cite{arcexp} are $^{158}$Gd,
$^{158}$Dy, $^{170}$Er, $^{178}$Hf, $^{236}$U, and $^{248}$Cm.

From the $\beta_0$ and $C$ values listed in Table 1, one can see
that the above condition is closely fulfilled. However, since the
ES-D solution is applicable mostly to nuclei with $R_{4/2}\geq
3.0$, the above mentioned condition describes only a small part of
the arc of regularity close to the SU(3) limit.

\section{Conclusions} 

In the present paper, an exactly separable version of the Bohr
Hamiltonian, called ES-D, which uses a potential of the form
$u(\beta)+u(\gamma)/\beta^2$, with a Davidson potential $\beta^2
+\beta_0^4/\beta^2$ in the place of $u(\beta)$, and a harmonic
oscillator with a minimum at $\gamma=0^{\circ}$ as $u(\gamma)$, is
developed. All bands (e.g., ground, $\beta$ and $\gamma$) in this
solution are treated on an equal footing, depending on two
parameters, the Davidson parameter $\beta_0$ and the stiffness $c$
of the $\gamma$-potential. The solution is found to be applicable
only to well deformed nuclei (with $R_{4/2}\geq 3.0$) due to the
$\beta^2$ denominator in the $u$($\gamma$) term. Nevertheless, it
reproduces very well the bandheads and energy spacings within
bands of almost all rare earth and actinide nuclei, with
$R_{4/2}\geq 3.0$,  for which available data exists, as well as
most of the inter-ground and intra-$\gamma$ band $B(E2)$
transition rates. The most glaring discrepancy concerns $B$($E$2)
values for the $\beta$ band to ground band transitions which are
typically overpredicted by an order of magnitude.  The two
exceptions where ES-D does not provide a good description of
energy spectra are $^{152}$Sm and $^{154}$Gd, which have
previously been shown to be well reproduced with the infinite
square well potential of the critical point symmetry X(5).
Furthermore, the ES-D solution provides insights regarding the
recently found correlation between the $\gamma$ stiffness and the
$\gamma$-bandhead energy, as well as the long standing problem of
producing a level scheme with IBA SU(3) degeneracies within the
framework of the Bohr Hamiltonian.

However, several open questions remain, in particular, concerning
the discrepancies in the $B$($E2$) predictions.  The
underprediction of the $\gamma \to $ ground and $\gamma \to \beta$
$B(E2)$s can be attributed to two reasons. First, the $\beta^2$
denominator in the $u(\gamma)$ term, ``pushes'' the nucleus to
more rigid axial behavior.  This can be investigated through a
detailed comparison of $B(E2)$s predicted by ES-X(5) and the exact
numerical solution of Ref. \cite{Caprio}, since the same
$u(\beta)$ and $u(\gamma)$ potentials are used in both cases. Work
in this direction is in progress. The second reason is the use of
a harmonic oscillator potential for $u(\gamma)$, as an
approximation valid for small $\gamma$, instead of a potential
periodic in $\gamma$. This can be studied through the use of a
periodic $\gamma$ potential \cite{Raduta} in ES-D, since no
approximations will be present in this case.

Furthermore, an exact numerical solution parallel to Ref.
\cite{Caprio} utilizing a $u(\beta)+u(\gamma)/\beta^2$ potential
with a Davidson potential as  $u(\beta)$ should demonstrate the
degree of importance of $\beta$-$\gamma$ coupling when compared to
the present results.

\section*{Acknowledgements}

Valuable discussions with F. Iachello, R.V. Jolos and M.A. Caprio
are acknowledged. This work was supported by U.S. DOE Grant No.
DE-FG02-91ER-40609.

\section*{Appendix I: Spectrum of ES-D}

One seeks \cite{IacX5} solutions of the relevant Schr\"odinger
equation having the form $ \Psi(\beta, \gamma, \theta_i)=
\phi_K^L(\beta,\gamma) {\cal D}_{M,K}^L(\theta_i)$, where
$\theta_i$ ($i=1$, 2, 3) are the Euler angles, ${\cal
D}(\theta_i)$ denote Wigner functions of them, $L$ are the
eigenvalues of angular momentum, while $M$ and $K$ are the
eigenvalues of the projections of angular momentum on the
laboratory-fixed $z$-axis and the body-fixed $z'$-axis
respectively.

As pointed out in Ref. \cite{IacX5}, in the case in which the
potential has a minimum around $\gamma =0$ one can write  the
angular momentum term of Eq. (\ref{eq:e1}) in the form
\begin{equation}\label{eq:e3}
\sum _{k=1,2,3} {Q_k^2 \over \sin^2 \left( \gamma -{2\pi \over 3}
k\right)} \approx {4\over 3} (Q_1^2+Q_2^2+Q_3^2) +Q_3^2 \left(
{1\over \sin^2\gamma} -{4\over 3}\right).
\end{equation}
Using this result in the Schr\"odinger equation corresponding to
the Hamiltonian of Eq. (\ref{eq:e1}), introducing \cite{IacX5}
reduced energies
 $\epsilon = 2B E /\hbar^2$ and reduced potentials $u = 2B V /\hbar^2$,
and assuming that the reduced potential can be separated into two
terms of the form  $u(\beta, \gamma) = u(\beta) +
u(\gamma)/\beta^2$, as in Refs. \cite{Wilets,Fort,Fort2,Heyde},
the Schr\"odinger equation can be separated into two equations
\begin{equation} \label{eq:e5}
\left[ -{1\over \beta^4} {\partial \over \partial \beta} \beta^4
{\partial \over \partial \beta} + {L(L+1)\over 3\beta^2} +u(\beta)
+ {\lambda\over \beta^2} \right] \xi_L(\beta) = \epsilon
\xi_L(\beta),
\end{equation}
\begin{equation}\label{eq:e6}
\left[ -{1\over \sin 3\gamma} {\partial \over
\partial \gamma}\sin 3\gamma {\partial \over \partial \gamma}
+{K^2 \over 4}  \left({1\over  \sin^2 \gamma}-{4\over 3}\right)
 +u(\gamma)\right] \eta_K(\gamma) =
\lambda \eta_K(\gamma).
\end{equation}

Eq. (\ref{eq:e6}) for $\gamma \approx 0$ can be treated as in Ref.
\cite{IacX5}, considering a potential of the form $u(\gamma)=
(3c)^2 \gamma^2$ and expanding in powers of $\gamma$. Then Eq.
(\ref{eq:e6}) takes the form
\begin{equation}\label{eq:e6a}
\left[ -{1\over \gamma} {\partial \over \partial \gamma} \gamma
{\partial \over \partial \gamma} + {K^2\over 4\gamma^2} + (3c)^2
\gamma^2  \right] \eta_K(\gamma) =  \epsilon_\gamma
\eta_K(\gamma),
\end{equation}
with $\epsilon_\gamma =\lambda +{K^2\over 3}$. The solution is
given in terms of Laguerre polynomials \cite{IacX5}
\begin{equation}
\epsilon_\gamma = (3C)(n_\gamma+1), \qquad C=2c, \qquad
n_\gamma=0,1,2,\ldots,
\end{equation}
\begin{equation}
n_\gamma=0,\quad K=0; \qquad n_\gamma=1, \quad K=\pm 2; \qquad
n_\gamma=2, \quad K=0,\pm 4; \qquad \ldots,
\end{equation}
\begin{equation}\label{eq:e21}
\eta_{n_\gamma,|K|}(\gamma)= C_{n_\gamma,|K|} \gamma^{|K/2|}
e^{-(3c)\gamma^2/2} L_{\tilde n}^{|K/2|} (3c\gamma^2), \qquad
\tilde n=(n_\gamma -|K/2|)/2.
\end{equation}

Eq. (\ref{eq:e5}) is then solved exactly for the case in which
$u(\beta)$ is a Davidson potential \cite{Dav,Elliott,Bahri}
\begin{equation}\label{eq:e7a}
u(\beta)=\beta^2 + {\beta_0^4 \over \beta^2},
\end{equation}
where $\beta_0$ denotes the position of the minimum of the
potential. In this case the eigenfunctions are \cite{Mosh1555}
\begin{equation}\label{eq:e11}
F_n^L(\beta)= \left[ {2 n!\over \Gamma \left(n+a+{5\over
2}\right)}\right]^{1/2} \beta^a L_n^{a+{3\over 2}}(\beta^2)
e^{-\beta^2/2},
\end{equation}
where $\Gamma(n)$ stands for the $\Gamma$-function, $L_n^a(z)$
denotes the Laguerre polynomials, and
\begin{equation}\label{eq:e12}
a= -{3\over 2}+\sqrt{ {L(L+1)-K^2\over 3}+{9\over 4} + \beta_0^4 +
3C(n_\gamma+1) },
\end{equation}
while the energy eigenvalues are
\begin{equation}\label{eq:e13a}
E_{n,L}= 2n+a+{5\over 2}= 2n+1 + \sqrt{ { L(L+1)-K^2\over 3}
+{9\over 4} + \beta_0^4 + 3C (n_\gamma+1) }, \qquad n=0,1,2,\ldots
\end{equation}
For $K=0$ one has $L=0$, 2, 4, \dots, while for $K\neq 0$ one
obtains $L=K$, $K+1$, $K+2$, \dots

In the above, $n$ is the usual oscillator quantum number. A formal
correspondence between the energy levels of X(5) and the present
solution can be established through the relation
\begin{equation}\label{eq:e14}
n=s-1.
\end{equation}
It should be remembered, however, that the origin of the two
quantum numbers is different, $s$ labelling the order of a zero of
a Bessel function and $n$ labelling the number of zeros of a
Laguerre polynomial. In the present notation, the ground state
band corresponds to $n=0$ ($s=1$). For the energy states the
notation $E_{s,L} = E_{n+1,L}$ of Ref. \cite{IacX5} will be kept.

The full wave function reads
\begin{equation}
\Psi(\beta,\gamma,\theta_i) = F^L_n(\beta)
\eta_{n_\gamma,|K|}(\gamma) {\cal D}^L_{MK}(\theta_i),
\end{equation}
and should be properly symmetrized \cite{EG}
\begin{equation}
\Psi(\beta,\gamma,\theta_i)= F_n^L(\beta)
\eta_{n_\gamma,|K|}(\gamma) \sqrt{ 2L+1\over 16\pi^2
(1+\delta_{K,0})} \left( {\cal D}^L_{M,K} +(-1)^L {\cal
D}^L_{M,-K}\right).
\end{equation}

It should be noticed at this point that Eq. (\ref{eq:e5}) for
$\lambda=0$ takes the form appearing in the framework of a X(5)
solution with the infinite well potential replaced by a Davidson
potential, called X(5)-D in the usual terminology. From the
expression for $\lambda$ given below Eq. (\ref{eq:e6a}) it is
clear that $\lambda=0$ is achieved for $K=0$ and
$\epsilon_\gamma=0$, i.e. $C=0$. It is therefore proved that the
numerical results of the ES-D solution for the $K=0$ bands (ground
state band and beta bands) will coincide with the corresponding
results of X(5)-D. This result should be considered as a numerical
coincidence, because $C=0$ is not acceptable in the framework of
ES-D, since the approximation of $\gamma$ being close to zero
collapses in this case. The lowest $R_{4/2}$ value within the
present model is obtained for $\beta_0=0$ and $C=0$, which
corresponds to the X(5)-$\beta^2$ solution \cite{BonX5}, giving
$R_{4/2}=2.646$.  Thus, while $\beta_0$ suggests a spherical
shape, the contribution from centrifugal term in the potential
results in a non-zero value for the average deformation.

\section*{Appendix II: $B(E2)$ values of ES-D}

B(E2) transition rates
\begin{equation}
B(E2; L K \to L'K')= {5\over 16 \pi} { |\langle L' K' || T^{(E2)}
|| L K \rangle|^2  \over 2L+1}
\end{equation}
can be calculated using the quadrupole operator \cite{IacX5}
\begin{equation}\label{eq:TE2}
T^{(E2)} = t \beta \left[ {\cal D}^{(2)}_{\mu,0} \cos\gamma
+{1\over \sqrt{2}} \left( {\cal D}^{(2)}_{\mu,2}+ {\cal
D}^{(2)}_{\mu,-2}\right) \sin\gamma \right],
\end{equation}
where $t$ is a scale factor, and the Wigner-Eckart theorem in the
form
\begin{equation}
\langle L' M' K' | T^{(E2)}_{\mu} | L M K \rangle = {1\over
\sqrt{2L'+1}} \langle L 2 L' | M \mu M'\rangle \langle L' K' ||
T^{(E2)} || LK\rangle .
\end{equation}
In  ground $\to$ ground, $\beta\to $ ground, $\beta \to \beta$ and
$\gamma \to \gamma$ transitions, only the first term of Eq.
(\ref{eq:TE2}) contributes, since the relevant angular momentum
coupling coefficients involving the second term vanish, while in
$\gamma \to $ ground and $\gamma \to \beta$ transitions only the
second term of Eq. (\ref{eq:TE2}) contributes, since the relevant
angular momentum coupling coefficients involving the first term
vanish. The final result reads
\begin{equation}
B(E2; n L n_\gamma K \to n' L' n'_\gamma K') = {5 \over 16\pi} t^2
(\langle L 2 L' | K, K'-K, K'\rangle)^2 B_{n,L,n',L'}^2
C_{n_\gamma,K,n'_\gamma,K'}^2,
\end{equation}
where
\begin{equation}
B_{n,L,n',L'} = \int \beta F_n^L(\beta) F_{n'}^{L'}(\beta) \beta^4
d\beta
\end{equation}
is the integral over $\beta$, while $C_{n_\gamma,K,n'_\gamma,K'}$
is the integral over $\gamma$, in agreement to Ref. \cite{Bijker}.
In  ground $\to $ ground, $\beta\to $ ground, $\beta \to \beta$
and $\gamma \to \gamma$ transitions ($\Delta K=0$ transitions),
the integral over $\gamma$ becomes $ C_{n_\gamma,K,n'_\gamma,K'} =
\delta_{n_\gamma, n'_\gamma} \delta_{K,K'}$, since (considering
$\cos\gamma \approx 1$) it corresponds to the relevant
orthonormality condition of the $\gamma$ wavefunctions, while in
$\gamma \to $ ground and $\gamma \to \beta$ transitions ($\Delta
K=2$ transitions) this integral has the form
\begin{equation}\label{eq:C}
 C_{n_\gamma,K,n'_\gamma,K'} = \int \sin\gamma \eta_{n_\gamma,|K|}
\eta_{n'_\gamma,|K'|} |\sin 3\gamma| d\gamma,
\end{equation}
since the volume element is \cite{Bohr}
\begin{equation}
d\tau = \beta^4 |\sin 3\gamma| \sin\theta d\beta d\gamma d\theta
d\phi d\psi .
\end{equation}
For the bands considered here one needs the special cases of Eq.
(\ref{eq:e21})
\begin{equation}
\eta_{0,0}= C_{0,0} e^{-(3c)\gamma^2/2}, \qquad \eta_{1,2}=
C_{1,2} \gamma e^{-(3c)\gamma^2/2},
\end{equation}
where the Laguerre polynomials are unity since $\tilde n=0$ in
both cases, as seen from Eq. (\ref{eq:e21}), the relevant
normalization conditions being
\begin{equation}\label{eq:CC}
(C_{0,0})^2 \int          e^{-(3c)\gamma^2} |\sin 3\gamma| d\gamma
=1,\qquad (C_{1,2})^2 \int \gamma^2 e^{-(3c)\gamma^2} |\sin
3\gamma| d\gamma =1.
\end{equation}
Then Eq. (\ref{eq:C}) takes the form
\begin{equation}
C_{1,2,0,0}= C_{0,0} C_{1,2} \int \gamma^2 e^{-(3c)\gamma^2} |\sin
3\gamma| d\gamma,
\end{equation}
in which the integral is the same as the one appearing in the
second normalization condition in Eq. (\ref{eq:CC}), resulting in
\begin{equation}
C_{1,2,0,0}= {C_{0,0}\over C_{1,2}}.
\end{equation}
Using the approximation $|\sin 3\gamma|\approx 3|\gamma|$ and the
integral
$$ \int_0^\infty x^m e^{-a x^2} dx = {\Gamma\left( {m+1\over 2}\right) \over
2 a^{m+1\over 2}} $$ the normalization conditions give
$$ (C_{0,0})^2= 2 c, \qquad (C_{1,2})^2 = 6 c^2, \qquad
{C_{0,0}\over C_{1,2}}= {1\over \sqrt{3c}}.$$ The normalization is
consistent with the one used by Bohr \cite{Bohr}. The same
approximations are also used in Ref. \cite{David}.

\section*{Appendix III: ES-X(5)}

In the case of the ES-X(5) solution \cite{ESX5}, in which
$u(\beta)$ is an infinite well potential
\begin{equation}
 u(\beta) = \left\{ \begin{array}{ll} 0 & \mbox{if $\beta \leq \beta_W$} \\
\infty  & \mbox{for $\beta > \beta_W$} \end{array} \right. ,
\end{equation}
the $\beta$-equation becomes a Bessel equation with energy
eigenvalues \cite{IacX5}
\begin{equation}
\epsilon_{\beta; s,L} = (k_{s,L})^2, \qquad k_{s,L}=  {x_{s,L}
\over \beta_W},
\end{equation}
where $x_{s,L}$ is the $s$-th zero of the Bessel function
$J_\nu(k_{s,L}\beta)$ with
\begin{equation}\label{eq:nu}
\nu = \sqrt{ {L(L+1)-K^2 \over 3} + {9\over 4} + 3C (n_\gamma+1)
},
\end{equation}
while the relevant eigenfunctions  are
\begin{equation}
\xi_{s,L}(\beta) = C_{s,L} \beta^{-3/2} J_\nu(k_{s,L} \beta),
\end{equation}
where $C_{s,L}$ are  normalization constants, determined from the
condition
\begin{equation}
\int_0^{\beta_W} \beta^4 \xi_{s,L}^2(\beta) d\beta =1,
\end{equation}
leading to
\begin{equation}
{1\over C_{s,L}^2} = {\beta_W^2 \over 2} J_{\nu+1}^2(x_{s,L}).
\end{equation}
The full wave function reads
\begin{equation}
\Psi(\beta,\gamma,\theta_i) = C_{s,L} \beta^{-3/2} J_\nu(k_{s,L}
\beta) \eta_{n_\gamma,|K|}(\gamma) {\cal D}^L_{MK}(\theta_i),
\end{equation}
and should be properly symmetrized \cite{EG}
\begin{equation}
\Psi(\beta,\gamma,\theta_i) = C_{s,L} \beta^{-3/2} J_\nu(k_{s,L}
\beta) \eta_{n_\gamma,|K|}(\gamma) \sqrt{ 2L+1\over 16\pi^2
(1+\delta_{K,0})} \left( {\cal D}^L_{M,K} +(-1)^L {\cal
D}^L_{M,-K}\right).
\end{equation}
In calculating $B(E2)$s, the integrals over $\gamma$ and the Euler
angles remain the same as in Appendix II, while the integrals over
$\beta$ take the form
\begin{equation}
B_{s,L,s',L'} =C_{s,L} C_{s',L'}\int \beta J_{\nu}(k_{s,L}\beta)
J_{\nu'}(k_{s',L'}\beta) \beta d\beta,
\end{equation}
where the formal correspondence $n=s-1$ holds.

It should be noticed at this point that for $C=0$ the numerical
results of ES-X(5) for the ground state band and the beta bands
coincide with the results of X(5), as it can be seen from Eq.
(\ref{eq:nu}). As discussed at the end of Appendix I, this should
be considered as a numerical coincidence, because $C=0$ is not
allowed in the X(5) framework, since it destroys the $\gamma
\approx 0$ approximation.

\newpage


\begin{figure}[ht]
\center{{\includegraphics[height=150mm]{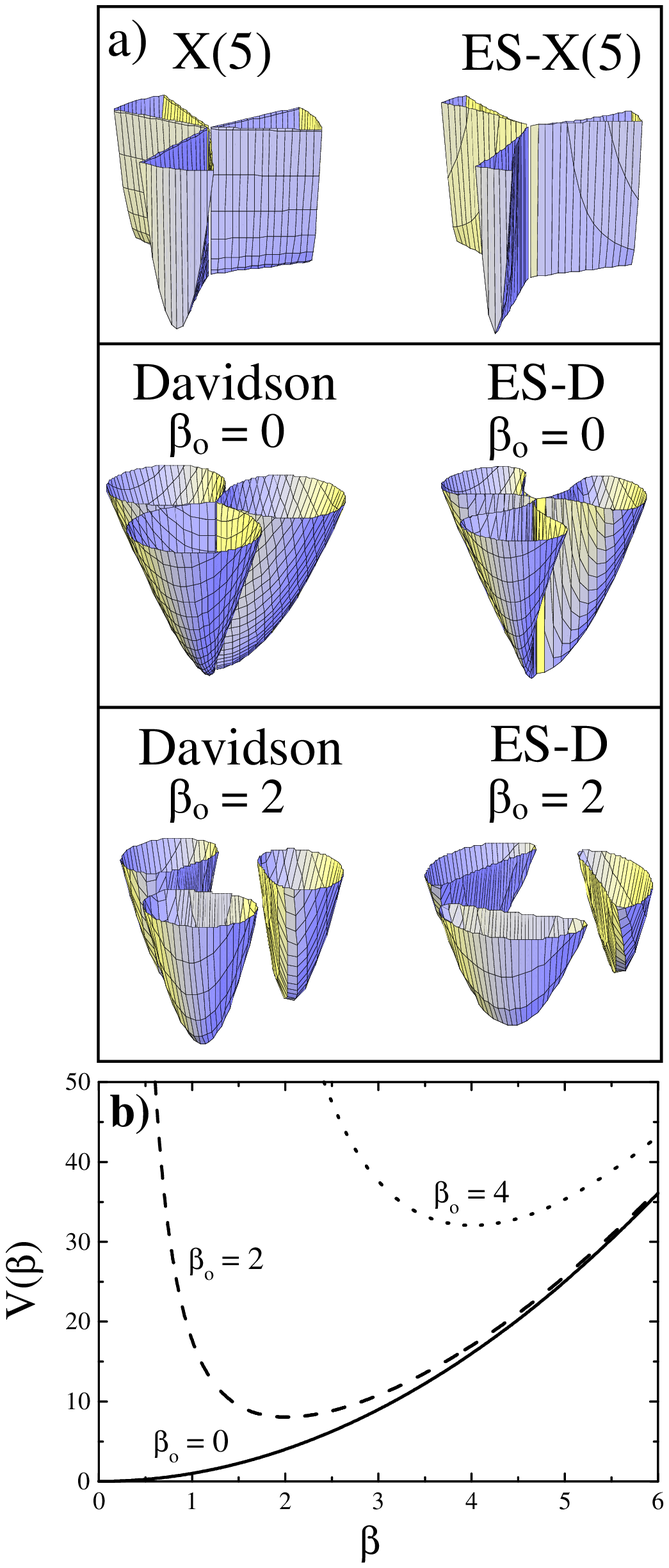}}}
\caption{(a) Potentials in both the $\beta$ and $\gamma$ degrees
of freedom for X(5) (top) and the Davidson potential with
$\beta_0$ = 0 (middle) and $\beta_0$ = 2 (bottom). Potentials are
shown for the approximate separation of variables (left) and the
exact separation of the variables (right). (b) Davidson potential
in the $\beta$ degree of freedom for a few values of the parameter
$\beta_0$.}
\end{figure}


\begin{figure}[ht]
\center{\includegraphics[height=150mm]{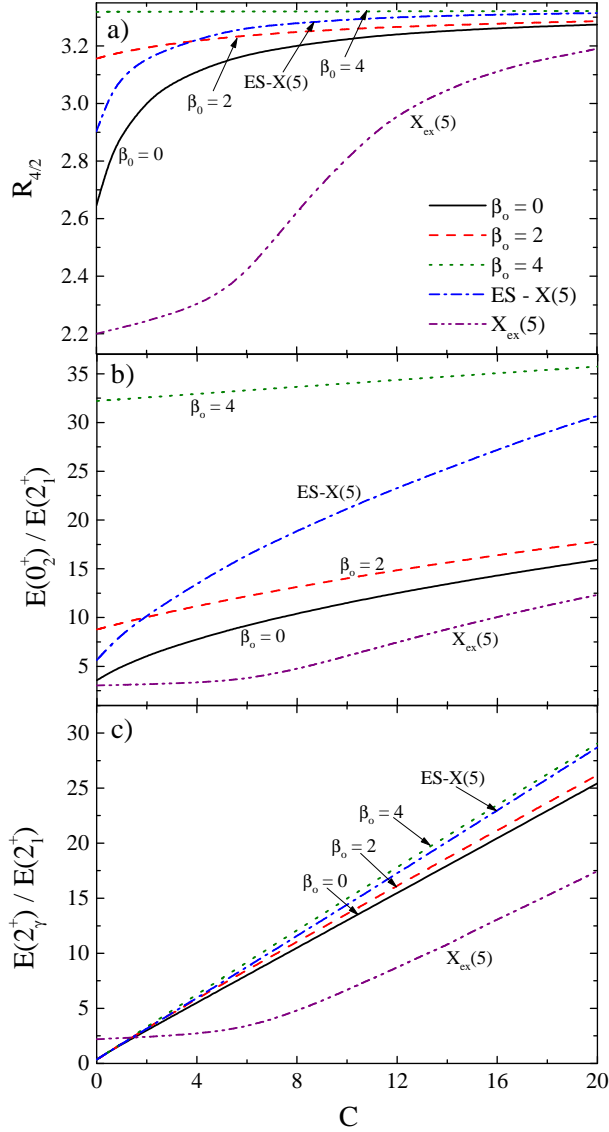}}
\caption{(Color online) (a) The $R_{4/2}=E(4_1^+)/E(2_1^+)$ ratio
as a function of the parameter $C$ for the ES-D solution (for
different values of the Davidson parameter $\beta_0$), the ES-X(5)
solution, and for the X$_{\textrm {ex}}$(5) solution
\cite{Caprio}. The X$_{\textrm {ex}}$(5) parameter $a$ is
connected to parameter $C$ of the present solution through the
relation $C = (2/3)\sqrt{a}$. (b) Same as (a), but for the ratio
$R_{0/2}=E(0_{\beta}^+)/E(2_1^+)$, corresponding to the normalized
$\beta$ bandhead energy. (c) Same as (a), but for the ratio
$R_{2/2}=E(2_{\gamma}^+)/E(2_1^+)$, corresponding to the
normalized $\gamma$ bandhead energy. See subsec. 3.1 for further
discussion.}
\end{figure}


\begin{figure}[ht]
\center{\includegraphics[height=180mm]{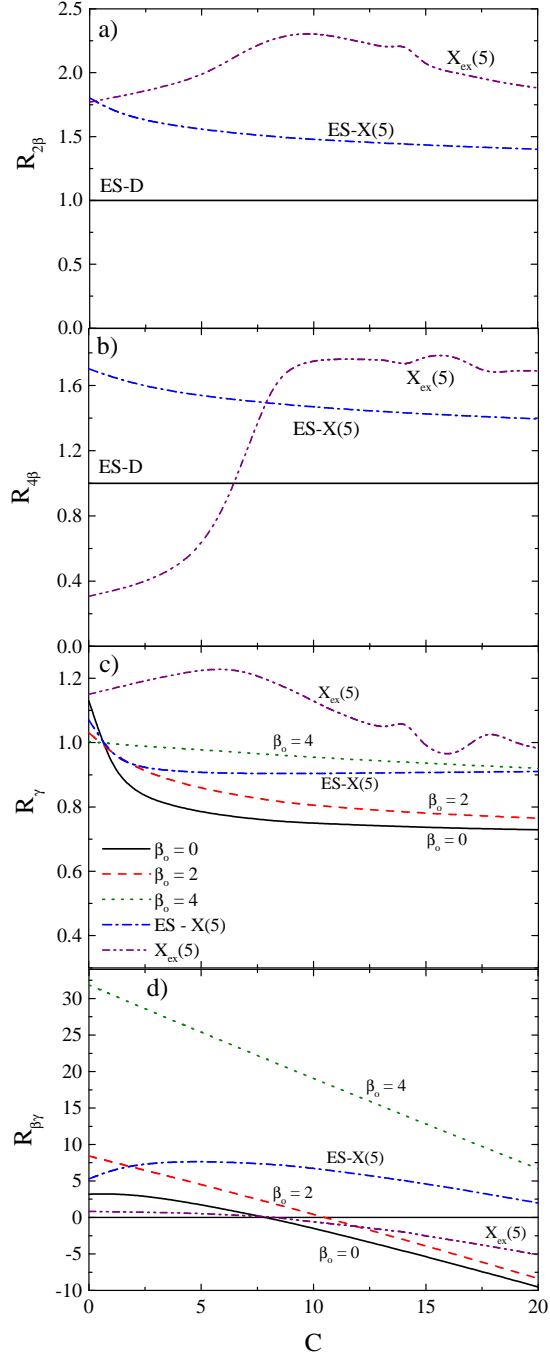}}
\caption{(Color online) Energy ratios (a)
$R_{2\beta}={E(2_{\beta}^+)-E(0_{\beta}^+) \over E(2_1^+)}$ [Eq.
(\ref{eq:e41})], (b) $R_{4\beta}={E(4_{\beta}^+)-E(2_{\beta}^+)
\over E(4_1^+)-E(2_1^+)}$ [Eq. (\ref{eq:e42})], (c)
$R_{\gamma}={E(4_{\gamma}^+)-E(2_{\gamma}^+) \over
E(4_1^+)-E(2_1^+)}$ [Eq. (\ref{eq:e43})], and (d)
$R_{\beta\gamma}= {E(0_{\beta}^+)-E(2_{\gamma}^+) \over E(2_1^+)}$
[Eq. (\ref{eq:e44})] as functions of the parameter $C$, for the
same solutions shown in Fig. 2. See subsec. 3.2 for further
discussion. }
\end{figure}


\begin{figure}[ht]
\center{{\includegraphics[height=100mm]{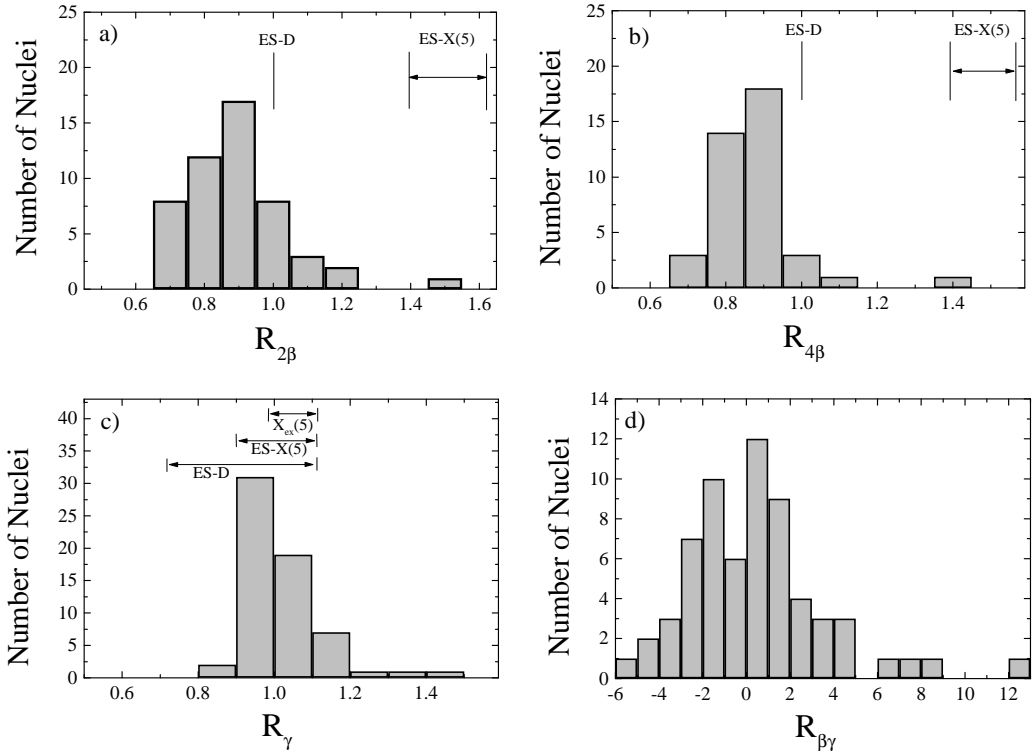}}} \caption{
Experimental data for the same energy ratios shown in Fig. 3~. For
each ratio, all nuclei with $A>100$ and $R_{4/2}$ $>$ 3.0 for
which sufficient experimental data (taken from Ref. \cite{NDS})
exist, have been taken into account.  The predictions for ES-D,
ES-X(5) and X$_{\textrm {ex}}$(5) are indicated in (a), (b), and
(c). The X$_{\textrm {ex}}$(5) predictions lie off scale to the
right in (a) and (b). See subsec. 3.2 for further discussion. }
\end{figure}


\begin{figure}[ht]
\center{{\includegraphics[height=200mm]{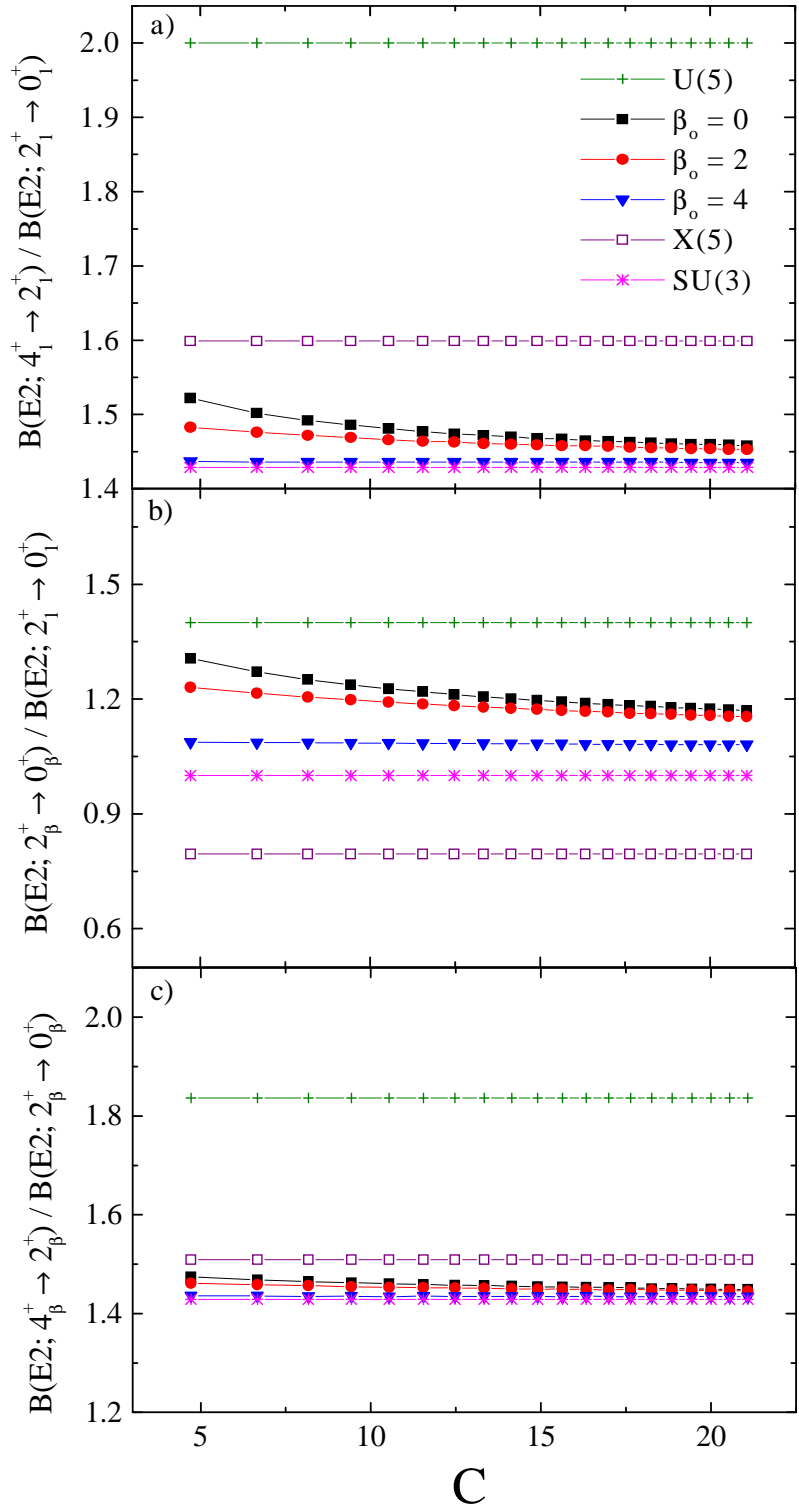}}}
\caption{(Color online)Intraband $B(E2)$ ratios for the ground
state and $\beta$ bands vs. the parameter $C$ as predicted by the
ES-D model (labelled by the value of the $\beta_0$ parameter),
 compared to U(5) \cite{IA}, X(5)
\cite{IacX5,Bijker}, and SU(3) \cite{IA} predictions, as described
in subsec. 3.3. }
\end{figure}


\begin{figure}[ht]
\center{{\includegraphics[height=200mm]{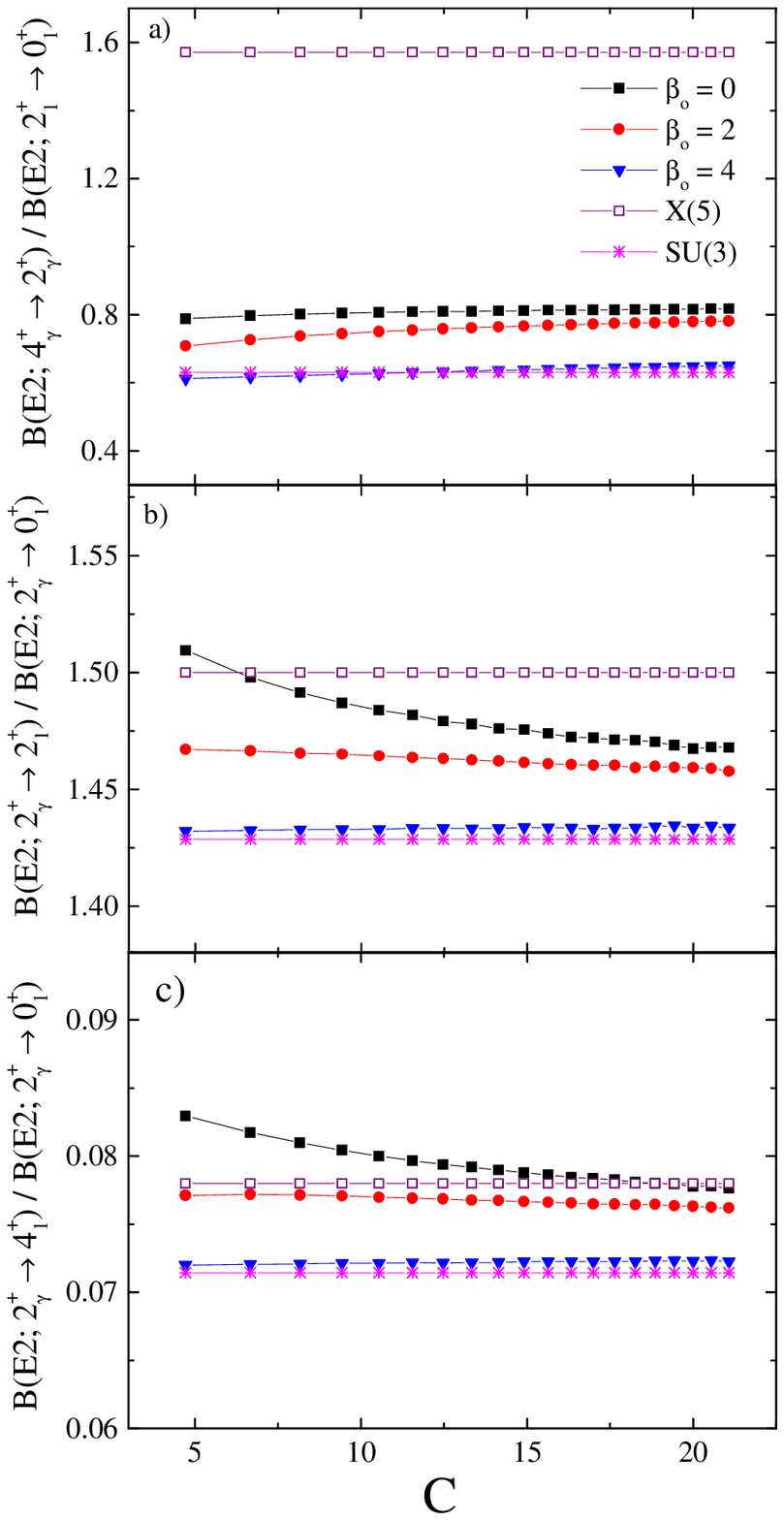}}}
\caption{(Color online) $B(E2)$ ratios from the $\gamma$ band vs.
the parameter $C$ as predicted by the ES-D model (labelled by the
value of the $\beta_0$ parameter), compared to the X(5)
\cite{IacX5,Bijker} and SU(3) \cite{IA} predictions, as described
in subsec. 3.3.}
\end{figure}


\begin{figure}[ht]
\center{{\includegraphics[height=200mm]{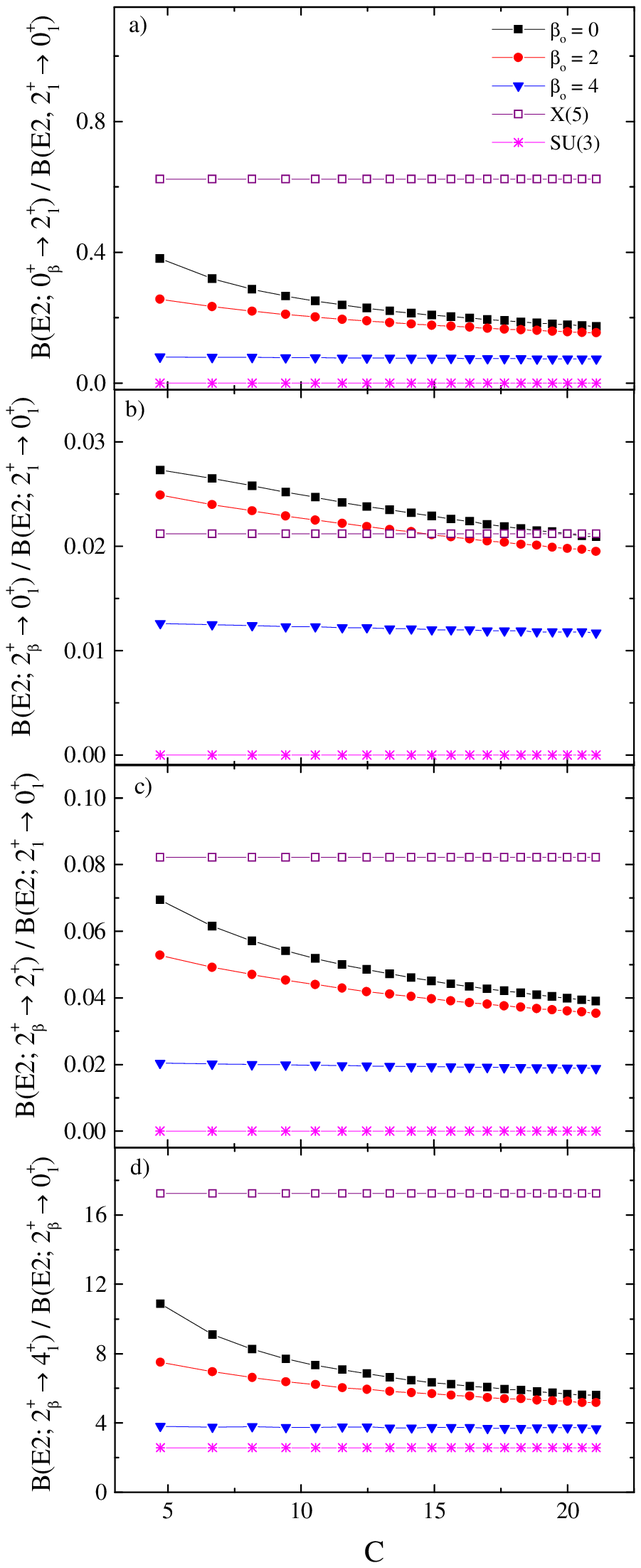}}}
\caption{(Color online) $B(E2)$ ratios from the $\beta$ band vs.
the parameter $C$ as predicted by the ES-D model (labelled by the
value of the $\beta_0$ parameter), compared to the X(5)
\cite{IacX5,Bijker} and SU(3) \cite{IA} predictions, as described
in subsec. 3.3. }
\end{figure}


\begin{figure}[ht]
\center{{\includegraphics[height=200mm]{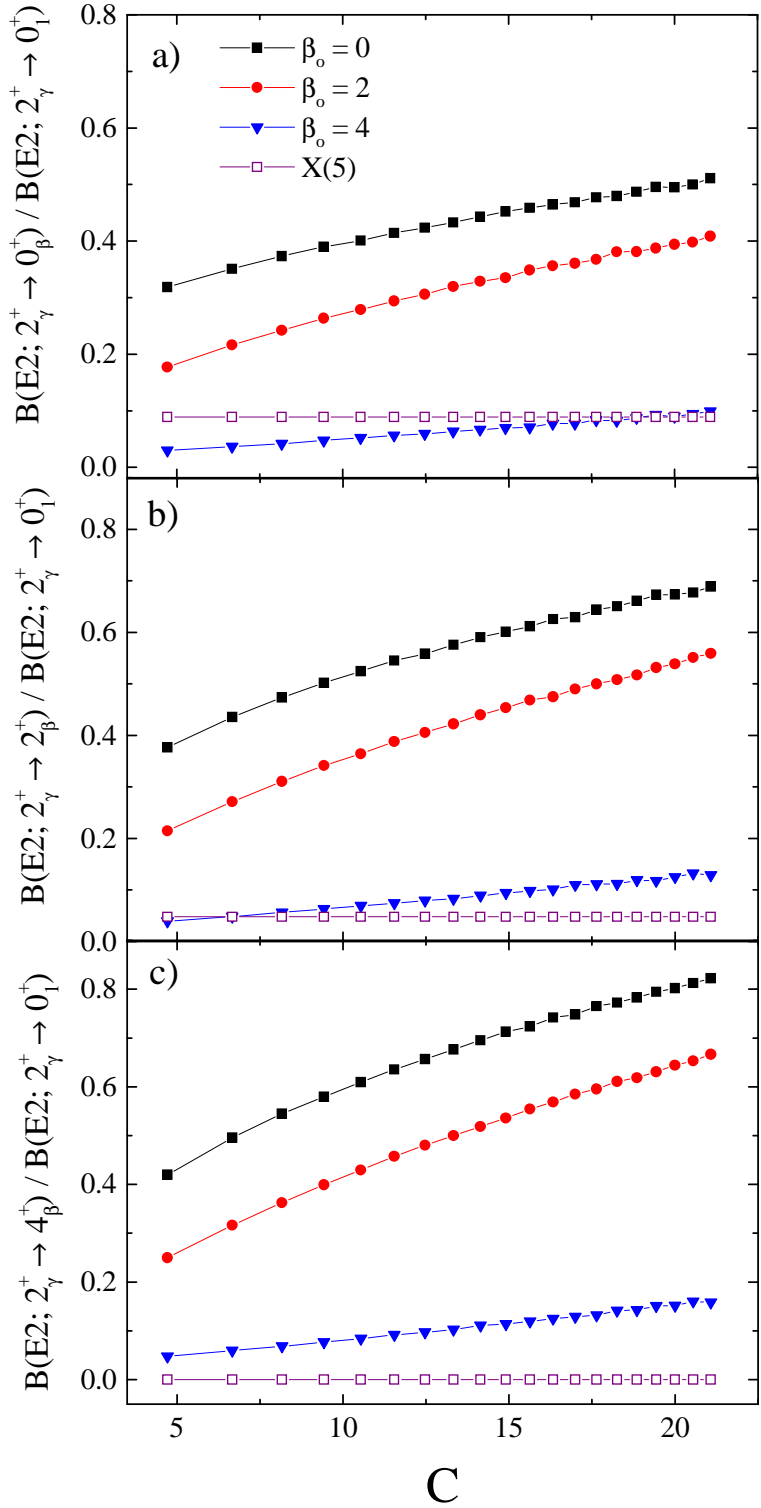}}}
\caption{(Color online) $B(E2)$ ratios between the $\gamma$ band
and $\beta$ band vs. the parameter $C$ as predicted by the ES-D
model (labelled by the value of the $\beta_0$ parameter), compared
to the X(5) \cite{IacX5,Bijker} predictions, as described in
subsec. 3.3. }
\end{figure}


\begin{figure}[ht]
\center{{\includegraphics[height=200mm]{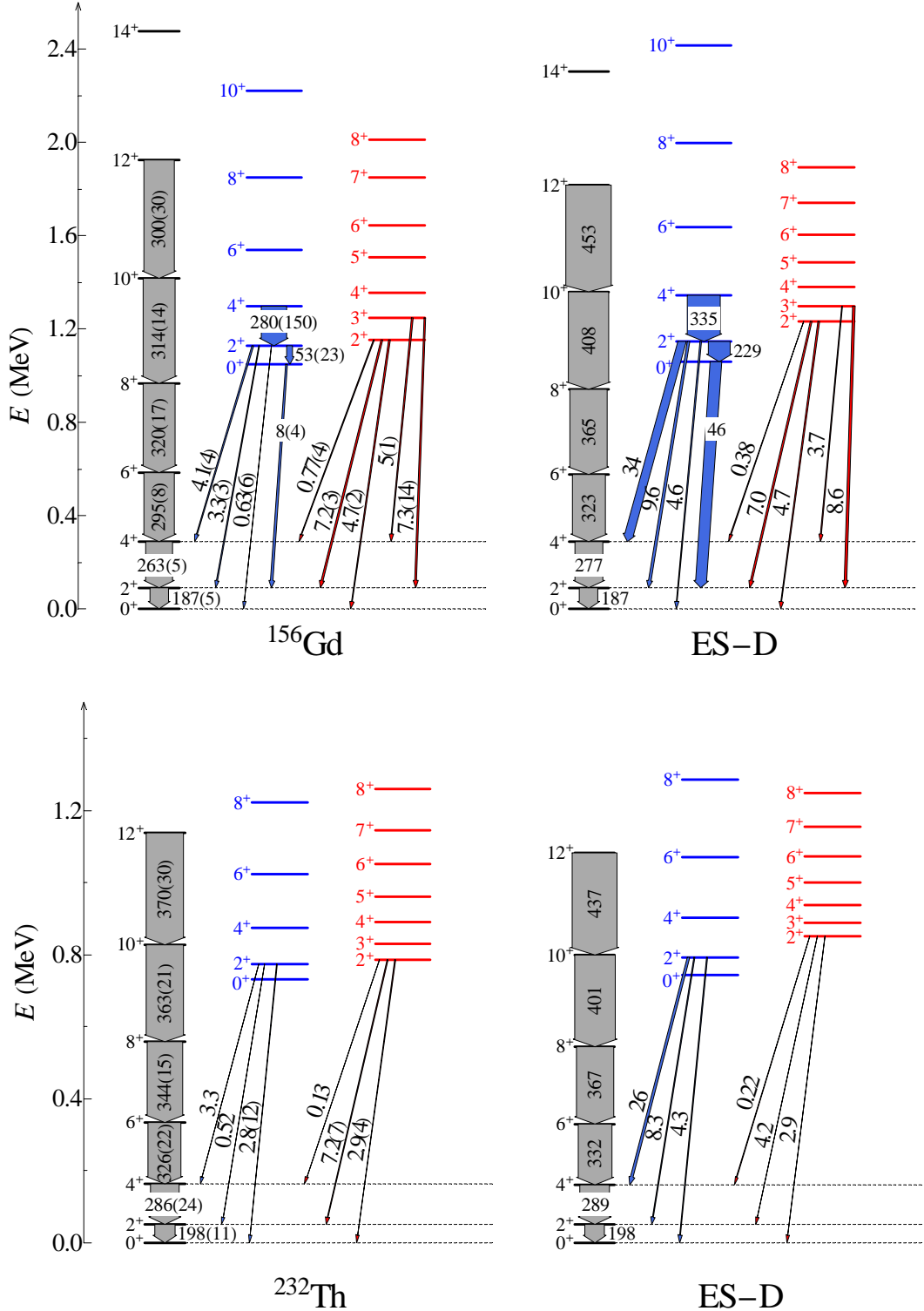}}}
\caption{(Color online) Experimental \cite{NDS} level schemes
(left) compared to ES-D predictions (right) for $^{156}$Gd (top)
and $^{232}$Th (bottom) using the parameter sets given in Table 1.
$\Delta K=0$ transitions are normalized to $2_1^+\to 0_1^+$, while
$\Delta K=2$ transitions are normalized to $2_\gamma^+\to 0_1^+$.}
\end{figure}


\begin{figure}[ht]
\center{{\includegraphics[height=100mm]{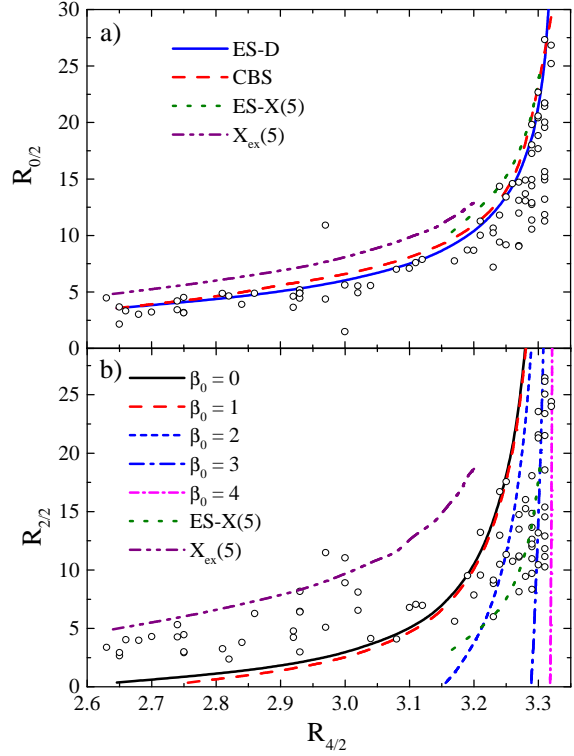}}}
\caption{(Color online) (a) Experimental data \cite{NDS} for the
normalized $\beta$ bandhead energies $R_{0/2} =
E(0_{\beta}^+)/E(2_1^+)$ vs. the energy ratio $R_{4/2} =
E(4_1^+)/E(2_1^+)$, compared to predictions of the the ES-D and
ES-X(5) solutions, as well as to the predictions of the confined
$\beta$-soft (CBS) solution \cite{CBS1,CBS2}. (b) Same as (a), but
for the normalized $\gamma$ bandheads $R_{2/2} =
E(2_{\gamma}^+)/E(2_1^+)$. See subsection 3.5 for further
discussion. }
\end{figure}


\begin{figure}[ht]
\center{{\includegraphics[height=100mm]{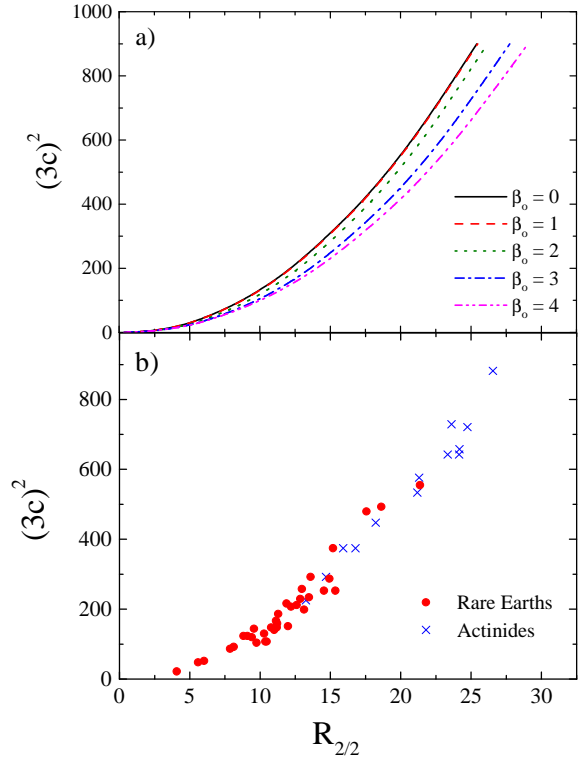}}}
\caption{(Color online) (a) The gamma-stiffness coefficient
$(3c)^2$ is shown as a function of the normalized $\gamma$
bandhead energy $R_{2/2}=E(2_{\gamma}^+)/E(2_1^+)$ predicted by
the ES-D solution for different values of the Davidson parameter
$\beta_0$. (b) Same as (a), but for the rare earth and actinide
nuclei appearing in Table 1~. See subsec. 3.6 for further
discussion. }
\end{figure}

\newpage

\begin{table}

\caption{Comparison of theoretical predictions of the exactly
separable Davidson solution [ES-D] to experimental data\cite{NDS}
of rare earth and actinides with $R_{4/2}$ $>$ 3.0 and known
$0_2^+$ and $2_{\gamma}^+$ states.  The
$R_{4/2}=E(4_1^+)/E(2_1^+)$ ratios, as well as the $\beta$ and
$\gamma$ bandheads, normalized to the $2_1^+$ state and labelled
by $R_{0/2}=E(0_{\beta}^+)/E(2_1^+)$ and
$R_{2/2}=E(2_{\gamma}^+)/E(2_1^+)$ respectively, are shown. The
angular momenta of the highest levels of the ground state, $\beta$
and $\gamma$ bands included in the rms fit are labelled by $L_g$,
$L_\beta$, and $L_\gamma$ respectively, while $n$ indicates the
total number of levels involved in the fit and $\sigma$ is the
quality measure of Eq. (\ref{eq:e99}). See subsec. 3.4 for further
discussion. }

\bigskip

\begin{tabular}{ l r r r  r r r r r r r r r r }
\hline nucleus & $R_{4/2}$ & $R_{4/2}$ & $R_{0/2}$& $R_{0/2}$
&$R_{2/2}$ & $R_{2/2}$ & $\beta_0$ & $C$   &
$L_g$ & $L_\beta$ & $L_\gamma$ & $n$ & $\sigma$ \\
        & exp &  th  & exp & th  & exp & th &  &  &  &  &  &   &  \\

\hline $^{154}$Sm & 3.25 & 3.26 & 13.4 & 14.1 & 17.6 & 18.8 & 1.26
& 14.6 &
 16 &   6 &   7 & 17 & 1.025 \\

$^{156}$Gd & 3.24 & 3.23 & 11.8 & 11.9 & 13.0 & 13.9 & 0.0  & 10.7
&
 14 & 10  &   8 & 19 & 1.044 \\
$^{158}$Gd & 3.29 & 3.27 & 15.0 & 14.8 & 14.9 & 15.3 & 2.05 & 11.3
& 12 &  6  &   6 & 14 & 0.624 \\
$^{160}$Gd & 3.30 & 3.29 & 17.6 & 18.3 & 13.1 & 13.4 & 2.69 & 9.4
& 16 &  4  &   8 & 17 & 0.962 \\
$^{162}$Gd & 3.29 & 3.30 & 19.8 & 20.2 & 12.0 & 12.0 & 2.93 & 8.2
& 14 &  0  &   4 & 10 & 0.335 \\

$^{158}$Dy & 3.21 & 3.20 & 10.0 & 10.4 &  9.6 & 10.5 &0.0  & 8.0 &
14 &  8 &  8 & 18 & 0.928 \\
$^{160}$Dy & 3.27 & 3.27 & 14.7 & 15.8 & 11.1 & 12.1 & 2.40 & 8.6
&28 & 4 & 23 & 38 & 0.633 \\
$^{162}$Dy & 3.29 & 3.29 & 17.3 & 17.7 & 11.0 & 11.4 &2.68 & 7.9 &
18 & 6 & 11 & 22 & 1.109 \\
$^{164}$Dy & 3.30 & 3.30 & 22.6 & 22.5 & 10.4 & 10.3 &3.18 & 6.9 &
20 & 0 & 10 & 19 & 0.089 \\
$^{166}$Dy & 3.31 & 3.27 & 15.0 & 14.9 & 11.2 & 11.4 & 2.30 & 8.1
&6 & 2 & 5 & 8 & 0.166 \\

$^{160}$Er & 3.10 & 3.15 & 7.1 & 8.7 & 6.8 & 7.0 & 0.0 & 5.2 & 14
& 2 & 5 & 12 & 0.815\\
$^{162}$Er & 3.23 & 3.23 & 10.7  & 11.6 &  8.8 & 10.1 & 1.68 & 7.4
&12 & 4 & 11 & 18 & 0.942 \\
$^{164}$Er & 3.28 & 3.26 & 13.6 & 13.9 &  9.4 & 10.3 & 2.19 & 7.3
&14 & 6 & 11 & 20 & 0.937 \\
$^{166}$Er & 3.29 & 3.28 & 18.1 & 17.3 &  9.8 & 9.9 &2.68 &  6.8 &
16 & 8 & 13 & 24 & 0.397 \\
$^{168}$Er & 3.31 & 3.27 & 15.3 & 15.0 & 10.3 & 10.8 & 2.34 & 7.6
& 12 & 6 & 8 & 16 & 0.892 \\
$^{170}$Er & 3.31 & 3.23 & 11.3 & 11.7 & 11.9 & 12.8 & 1.09 & 9.8
&10 & 4 & 7 & 13 & 0.978 \\

$^{164}$Yb & 3.13 & 3.20 & 7.9 & 10.1 & 7.0 & 7.5 & 1.58 &  5.4 &
18 & 0 & 5 & 13 & 0.807\\
$^{166}$Yb & 3.23& 3.20 & 10.2 & 10.5&  9.1 & 9.8 &1.18 &  7.4 &
12 & 6 & 7 & 15 & 0.774 \\
$^{168}$Yb & 3.27 & 3.25 & 13.2& 13.5  & 11.2 &11.6 & 2.03 & 8.4 &
14 & 4 & 7 & 15 & 0.532 \\
$^{170}$Yb & 3.29 & 3.25& 12.7 & 13.2 & 13.6 &15.0 & 1.56 & 11.4 &
12 & 6 & 11 & 19 & 1.168 \\
$^{172}$Yb & 3.31 & 3.26 & 13.2 & 13.8& 18.6 & 19.0 &0.0  & 14.8 &
12 & 8 & 5 & 14 & 1.078 \\
$^{174}$Yb & 3.31 & 3.30 & 19.4 & 20.0& 21.4 & 21.7 &2.66 & 15.7 &
16 & 4 & 5 & 14 & 0.956 \\
$^{176}$Yb & 3.31 & 3.30& 21.7 & 22.4& 15.4 & 15.4 &3.08 & 10.6 &
18 & 0 & 2 & 10 & 0.386 \\
$^{178}$Yb & 3.31 & 3.27& 15.7 & 15.5& 14.5 & 14.6 &2.24 & 10.6 &
6 & 4 & 2 & 6 & 0.128 \\

$^{168}$Hf & 3.11 & 3.16 & 7.6 & 9.0 & 7.1 & 7.5 & 0.0 & 5.6 & 14
& 2 & 4 & 11 & 0.814\\
$^{170}$Hf & 3.19 & 3.20 & 8.7 & 10.2 & 9.5 & 10.0 & 0.0 & 7.6 &
14 & 2 & 4 & 11& 0.835\\
$^{172}$Hf & 3.25 & 3.22 &  9.2 & 11.1 & 11.3 & 11.9 &0.0  & 9.1 &
12 & 2 & 6 & 12 & 1.142 \\
$^{174}$Hf & 3.27 & 3.23 &  9.1 & 11.6 & 13.5 & 13.2 &0.0  & 10.2
&10 & 2 & 4 & 9 & 1.471 \\
$^{176}$Hf & 3.28 & 3.25 & 13.0 & 13.1& 15.2 & 16.6 &0.99 & 12.9 &
12 & 6 & 8 & 16 & 1.040 \\
$^{178}$Hf & 3.29 & 3.25 & 12.9 & 13.3 & 12.6 & 13.1 &1.86 & 9.7 &
14 & 6 & 6 & 15 & 1.099 \\
$^{180}$Hf & 3.31 & 3.23 & 11.8 & 12.0& 12.9 & 13.2 &1.26 & 10.1 &
10 & 4 & 5 & 11 & 0.830 \\

\hline
\end{tabular}
\end{table}

\begin{table}
\setcounter{table}{0} \caption{(continued)}

\bigskip

\begin{tabular}{ l r r r  r r r r r r r r r r }
\hline nucleus & $R_{4/2}$ & $R_{4/2}$ & $R_{0/2}$& $R_{0/2}$
&$R_{2/2}$ & $R_{2/2}$ & $\beta_0$ & $C$   &
$L_g$ & $L_\beta$ & $L_\gamma$ & $n$ & $\sigma$ \\
        & exp &  th  & exp & th  & exp & th &  &  &  &  &  &   &  \\

\hline

$^{176}$W & 3.22 & 3.20 & 7.8 & 10.1 & 9.6 & 9.9 & 0.0 & 7.5 & 12
& 2 & 5 & 11 & 1.281\\
$^{178}$W  & 3.24 & 3.19&  9.4 & 9.8& 10.5 & 9.1& 0.0  & 6.9 &
 12 & 8 & 2 & 11 & 1.260 \\
$^{180}$W  & 3.26 & 3.27& 14.6& 14.7 & 10.8 & 11.4 &2.27 &  8.1 &
12 & 0 & 7 & 12 & 0.313 \\
$^{182}$W  & 3.29 & 3.24 & 11.3& 12.4 & 12.2 & 12.8 &1.60 &  9.6 &
12 & 4 & 6 & 13 & 1.136 \\
$^{184}$W  & 3.27 & 3.18 &  9.0 & 9.7&  8.1 & 8.6 & 0.98 &  6.4 &
 10 & 4 & 6 & 12 & 0.928 \\
$^{186}$W  & 3.23 & 3.14&  7.2 & 8.4 &  6.0 & 6.5 &0.0  &  4.8 &
10 & 4 & 6 & 12 & 1.142 \\

$^{180}$Os & 3.09 & 3.15 & 5.6 & 8.6 & 6.6 & 6.9 & 0.0 & 5.1 & 14
& 6 & 7 & 16 & 1.348\\
$^{184}$Os & 3.20 & 3.21 &  8.7 & 10.6 &  7.9 & 8.5 &1.58 & 6.2 &
12 & 0 & 6 & 11 & 0.918 \\
$^{186}$Os & 3.17 & 3.14 &  7.7 & 8.3 &  5.6 & 6.3 &0.0  &  4.6 &
10 & 10 & 9 & 18 & 0.982 \\
$^{188}$Os & 3.08 & 3.13 &  7.0 & 8.1 &  4.1 & 4.5 &1.42 &  3.1 &
12 & 2 & 7 & 13 & 0.571 \\

$^{228}$Ra & 3.21 & 3.23 & 11.3 & 11.5 & 13.3 & 13.0 & 0.0
& 10.0 & 6 & 4 & 3 & 7 & 0.447 \\

$^{228}$Th & 3.24 & 3.26& 14.4 & 14.5 & 16.8 & 17.1 &1.79 & 12.9 &
18 & 2 & 5 & 14 & 0.240 \\
$^{230}$Th & 3.27 & 3.24 & 11.9 & 12.3& 14.7 &14.7 & 0.0  & 11.4 &
12 & 4 & 4 & 11 & 0.864 \\
$^{232}$Th & 3.28 & 3.27 & 14.8 & 15.0 & 15.9 & 17.2 & 1.95 & 12.9
& 14 & 20 & 12 & 28 & 1.030 \\

$^{232}$U  & 3.29 & 3.26 & 14.5 & 14.6 & 18.2 & 18.5 &1.67 & 14.1
&14 & 10 & 4 & 15 & 0.910 \\
$^{234}$U  & 3.30 & 3.29 & 18.6 & 19.0 & 21.3 & 21.8 &2.50 & 16.0
& 18 & 8 & 7 & 19 & 0.634 \\
$^{236}$U  & 3.30 & 3.30 & 20.3& 20.8 & 21.2 & 21.4 &2.77 & 15.4 &
18 & 4 & 5 & 15 & 0.686 \\
$^{238}$U  & 3.30 & 3.30 & 20.6 & 21.7 & 23.6 & 24.8 &2.79 & 18.0
& 18 & 4 & 15 & 25 & 0.845 \\

$^{238}$Pu & 3.31 & 3.30 & 21.4 & 22.0 & 23.3 & 23.5 &2.86 & 16.9
&16 & 2 & 4 & 12 & 0.839 \\
$^{240}$Pu & 3.31 & 3.30 & 20.1 & 20.5 & 26.6 & 26.8 &2.56 & 19.8
& 16 & 4 & 4 & 13 & 0.878 \\
$^{242}$Pu & 3.31 & 3.30 & 21.5& 21.9 & 24.7 & 24.7 &2.82 & 17.9 &
16 & 2 & 2 & 10 & 0.740 \\

$^{248}$Cm & 3.31 & 3.31& 25.0 & 25.4& 24.2 & 24.2& 3.21 & 17.1
&20 & 4 & 2 & 13 & 0.520 \\

$^{250}$Cf & 3.32 & 3.31& 27.0 & 26.9& 24.2 & 24.1& 3.36 & 16.9
&8 & 2 & 4 & 8 & 0.067 \\

\hline
\end{tabular}
\end{table}

\begin{table}

\caption{Comparison of several $B(E2)$ ratios predicted (lower
line) by the exactly separable Davidson solution [ES-D], for the
parameter values shown in Table 1, to experimental data\cite{NDS}
(upper line) of several nuclei where the relevant data are known.
See subsec. 3.4 for further discussion. }

\bigskip

\begin{tabular}{l r@{.}l r@{.}l r@{.}l r@{.}l r@{.}l r@{.}l r@{.}l r@{.}l r@{.}l}

\hline
   \multicolumn{1}{l}{nucleus}
   &\multicolumn{2}{c} {$4_1\to 2_1 \over 2_1\to 0_1$}
    &\multicolumn{2}{c} {$6_1\to 4_1 \over 2_1\to 0_1$}
    &\multicolumn{2}{c} {$8_1\to 6_1 \over 2_1\to 0_1$}
   &\multicolumn{2}{c} {$10_1\to 8_1 \over 2_1\to 0_1$}
    &\multicolumn{2}{c} {$2_\beta \to 0_1 \over 2_1\to 0_1$}
   &\multicolumn{2}{c}{$2_\beta \to 2_1 \over 2_1\to 0_1$}
   &\multicolumn{2}{c}{$2_\beta \to 4_1 \over 2_1\to 0_1$}
   &\multicolumn{2}{c}{$2_\gamma\to 2_1 \over 2_\gamma \to 0_1$}
   &\multicolumn{2}{c}{$2_\gamma\to 4_1 \over 2_\gamma \to 0_1$} \\

   & \omit\span & \omit\span & \omit\span & \omit\span &
  \multicolumn{2}{c} {x $10^3$} &  \multicolumn{2}{c} {x $10^3$} &  \multicolumn{2}{c} {x
   $10^3$}
   & \omit\span & \omit\span \\

\hline $^{154}$Sm   & 1&40(5) & 1&67(7) & 1&83(11) & 1&81(11) &
5&4(13) &  \omit\span      & \multicolumn{2}{c} {25(6)} &
\omit\span
      & 0&21(5) \\
           & 1&47 & 1&69 & 1&88 & 2&06 & 22&7 & 44&5 &  \multicolumn{2}{c}  {142} &
1&47 & 0&08 \\

$^{156}$Gd & 1&41(5) & 1&58(6) & 1&71(10) & 1&68(9) &  3&4(3) &
\multicolumn{2}{c} {18(2)} & \multicolumn{2}{c} {22(2)} &
1&55(7) & 0&16(1) \\
           & 1&48 & 1&73 & 1&95 & 2&18 & 24&6 & \multicolumn{2}{c} {52} & \multicolumn{2}{c} {179} &
1&48 & 0&08 \\

$^{158}$Gd & 1&46(5) &  \omit\span        & 1&67(16) & 1&72(16) &
1&6(2) & 0&4(1) & 7&0(8) &
1&77(26) & 0&079(14) \\
           & 1&46 & 1&68 & 1&86 & 2&03 & 22&1 & 42&5 & \multicolumn{2}{c} {133} &
1&46 & 0&077 \\

$^{158}$Dy & 1&45(10) & 1&86(12) & 1&86(38) & 1&75(28) &
\multicolumn{2}{c} {12(3)} & \multicolumn{2}{l} {19(4)} &
\multicolumn{2}{c} {66(16)} &
3&22(94) & 0&36(15) \\
           & 1&49 & 1&77 & 2&02 & 2&29 & \multicolumn{2}{c} {26} & \multicolumn{2}{l} {58} & \multicolumn{2}{c} {215} &
1&49 & 0&08 \\

$^{160}$Dy & 1&46(7) & 1&23(7) & 1&70(16) & 1&69(9) &  3&4(4) &
\omit\span   & 8&5(10) &
1&89(18) & 0&13(1) \\
           & 1&46 & 1&67 & 1&84 & 2&00 & 21&3 & 40&2 & \multicolumn{2}{c} {122} &
1&45 & 0&08 \\

$^{162}$Dy & 1&45(7) & 1&51(10) & 1&74(10) & 1&76(13) & \omit\span
&  \omit\span   &  \omit\span   &
1&67(20) & 0&14(1) \\
           & 1&45 & 1&65 & 1&80 & 1&94 & 19&8 & 36&2 & \multicolumn{2}{c} {104} &
1&45 & 0&07 \\

$^{164}$Dy & 1&30(7) & 1&56(7) & 1&48(9) & 1&69(9) &  \omit\span &
\omit\span &  \omit\span   &
2&00(30) & 0&24(3) \\
           & 1&45 & 1&62 & 1&75 & 1&86 & 16&9 & 29&2 & \multicolumn{2}{c} {77} &
1&44 & 0&07 \\

$^{162}$Er &  \omit\span         &  \omit\span                &
\omit\span  & \omit\span     & \multicolumn{2}{c} {8(7)} &
\omit\span  & \multicolumn{2}{c} {170(90)} &
2&37(25) & 0&29(21) \\
           & 1&48 & 1&74 & 1&97 & 2&20 & 24&9 & 52&8 & \multicolumn{2}{c} {190} &
1&48 & 0&08 \\

$^{164}$Er & 1&18(13) &   \omit\span        & 1&57(9) & 1&64(11) &
\omit\span &  \omit\span  &  \omit\span   &
2&19(35) & 0&33(5) \\
           & 1&47 & 1&69 & 1&88 & 2&07 & 22&9 & 45&0 & \multicolumn{2}{c} {145} &
1&46 & 0&08 \\

$^{166}$Er & 1&45(12) & 1&62(22) & 1&71(25) & 1&73(23) &
\omit\span    &  \omit\span   &  \omit\span   &
1&76(18) & 0&12(1) \\
           & 1&46 & 1&65 & 1&81 & 1&95 & 20&1 & 36&9 & \multicolumn{2}{c} {107} &
1&45 & 0&07 \\

$^{168}$Er & 1&54(7) & 2&13(16) & 1&69(11) & 1&46(11) & \omit\span
&  \omit\span   &  \omit\span   &
1&77(10) & 0&129(9) \\
           & 1&46 & 1&68 & 1&85 & 2&03 & 21&9 & 42&1 & \multicolumn{2}{c} {131} &
1&46 & 0&076 \\

$^{170}$Er &  \omit\span         &  \omit\span         & 1&78(15)
& 1&54(11) & 1&4(1) & 0&2(2) & 6&8(12) &
 \omit\span        & 0&079(19) \\
           & 1&48 & 1&73 & 1&96 & 2&20 & 24&8 & 52&3 & \multicolumn{2}{c} {184} &
1&48 & 0&080 \\

$^{166}$Yb & 1&43(9) & 1&53(10) & 1&70(18) & 1&61(80) & \omit\span
&  \omit\span   &  \omit\span   &
     \omit\span    &  \omit\span          \\
           & 1&49 & 1&77 & 2&02 & 2&29 &  25&8 & 57&4 & \multicolumn{2}{c} {215} &
1&49 & 0&081 \\

$^{168}$Yb &  \omit\span         &   \omit\span             &
\omit\span  & \omit\span       & 8&6(9) &  \omit\span   &
\omit\span   &
2&09(50) & 0&39(10) \\
           & 1&47 & 1&70 & 1&89 & 2&09 & 23&2 & 46&2 & \multicolumn{2}{c} {151} &
1&47 & 0&08 \\

$^{170}$Yb &  \omit\span         &  \omit\span        & 1&79(16) &
1&77(14) & 5&4(10) &  \omit\span   &  \omit\span   &
1&78(50) & 0&18(5) \\
           & 1&47 & 1&70 & 1&90 & 2&11 & 23&5 & 47&1 & \multicolumn{2}{c} {156} &
1&47 & 0&08 \\

$^{172}$Yb & 1&42(10) & 1&51(14) & 1&89(19) & 1&77(11) & 1&1(1) &
3&7(6) & \multicolumn{2}{c} {12(1)} &
    \omit\span     & 0&097(11) \\
           & 1&47 & 1&69 & 1&88 & 2&08 & 22&9 & 45&2 & \multicolumn{2}{c} {146} &
1&48 & 0&079 \\

$^{174}$Yb & 1&39(7) & 1&84(26) & 1&93(12) & 1&67(12) & \omit\span
&  \omit\span  &  \omit\span  &
    \omit\span     & \omit\span          \\
           & 1&45 & 1&64 & 1&77 & 1&89 & 18&3 & 32&4 & \multicolumn{2}{c} {89} &
1&45 & 0&075 \\

$^{176}$Yb & 1&49(15) & 1&63(14) & 1&65(28) & 1&76(18) &
\omit\span    &  \omit\span   &  \omit\span   &
1&58(11) &  \omit\span          \\
           & 1&45 & 1&62 & 1&75 & 1&86 & 17&0 & 29&3 & \multicolumn{2}{c} {77} &
1&44 & 0&073 \\
\hline
\end{tabular}
\end{table}

\begin{table}
\setcounter{table}{1} \caption{ (continued) }

\bigskip

\begin{tabular}{l r@{.}l r@{.}l r@{.}l r@{.}l r@{.}l r@{.}l r@{.}l r@{.}l r@{.}l}

\hline
   \multicolumn{1}{l}{nucleus}
   &\multicolumn{2}{c} {$4_1\to 2_1 \over 2_1\to 0_1$}
    &\multicolumn{2}{c} {$6_1\to 4_1 \over 2_1\to 0_1$}
    &\multicolumn{2}{c} {$8_1\to 6_1 \over 2_1\to 0_1$}
   &\multicolumn{2}{c} {$10_1\to 8_1 \over 2_1\to 0_1$}
    &\multicolumn{2}{c} {$2_\beta \to 0_1 \over 2_1\to 0_1$}
   &\multicolumn{2}{c}{$2_\beta \to 2_1 \over 2_1\to 0_1$}
   &\multicolumn{2}{c}{$2_\beta \to 4_1 \over 2_1\to 0_1$}
   &\multicolumn{2}{c}{$2_\gamma\to 2_1 \over 2_\gamma \to 0_1$}
   &\multicolumn{2}{c}{$2_\gamma\to 4_1 \over 2_\gamma \to 0_1$} \\

   & \omit\span & \omit\span & \omit\span & \omit\span &
  \multicolumn{2}{c} {x $10^3$} &  \multicolumn{2}{c} {x $10^3$} &  \multicolumn{2}{c} {x
   $10^3$}
   & \omit\span & \omit\span \\

\hline $^{174}$Hf &  \omit\span            &   \omit\span       &
\omit\span &  \omit\span  & \multicolumn{2}{c}{14(4)} & \omit\span
& \multicolumn{2}{c} {9(3)} &
1&54(76) & \omit\span        \\
           & 1&48 & 1&74 & 1&96 & 2&20 & 24&8 & 52&5 & \multicolumn{2}{c}{185} &
1&49 & 0&0801 \\
$^{176}$Hf & \omit\span  & \omit\span      & \omit\span       &
\omit\span & 5&4(11) & \omit\span & \multicolumn{2}{c} {31(6)} &
\omit\span
      &   \omit\span     \\
           & 1&47 & 1&71 & 1&91 & 2&11 & 23&5 & 47&4 & \multicolumn{2}{c} {157} &
1&48 & 0&0791 \\
$^{178}$Hf & \omit\span       & 1&38(9) & 1&49(6) & 1&62(7) &
0&4(2) & \omit\span & 2&4(9) &
1&13(17) & 0&066(10) \\
           & 1&47 & 1&70 & 1&90 & 2&10 & 23&3 & 46&7 & \multicolumn{2}{c} {153} &
1&47 & 0&078 \\

$^{180}$Hf & 1&48(20) & 1&41(15) & 1&61(26) & 1&55(10) &
\omit\span & \omit\span & \omit\span &
1&34(28) & \omit\span        \\
           & 1&48 & 1&73 & 1&95 & 2&17 & 24&5 & 51&1 & \multicolumn{2}{c} {177} &
1&48 & 0&0795 \\

$^{182}$W  & 1&43(8) & 1&46(16) & 1&53(14) & 1&48(14) &  6&6(6) &
4&6(6) & \multicolumn{2}{c} {13(1)} &
1&98(7) & 0&010(1) \\
           & 1&48 & 1&72 & 1&93 & 2&15 & 24&1 & 49&6 & \multicolumn{2}{c} {169} &
1&48 & 0&0787 \\

$^{184}$W  & 1&35(12) & 1&54(9) & 2&00(18) & 2&45(51) & 1&8(3) &
\omit\span & \multicolumn{2}{c} {24(3)} &
1&91(13) & 0&109(9) \\
           & 1&50 & 1&79 & 2&07 & 2&37 &  26&5 & 61&4 & \multicolumn{2}{c} {240} &
1&50 & 0&0814 \\
$^{186}$W  & 1&30(9) & 1&69(12) & 1&60(12) & 1&36(36) & \omit\span
& \omit\span & \omit\span &
2&18(15) & \omit\span        \\
           & 1&52 & 1&85 & 2&18 & 2&53 & 27&3 & 69&0 & \multicolumn{2}{c} {294} &
1&51 & 0&0829 \\
$^{186}$Os & 1&45(7) & 1&99(7) & 1&89(11) & 2&06(44) & \omit\span
& \omit\span & \omit\span &
2&33(12) & 0&12(4) \\
           & 1&52 & 1&86 & 2&19 & 2&55 & 27&3 & 70&0 & \multicolumn{2}{c} {301} &
1&51 & 0&0830 \\
$^{188}$Os & 1&68(11) & 1&75(11) & 2&04(15) & 2&38(32)        &
\omit\span & \omit\span & \omit\span &
3&20(6) & 6&8(13) \\
           & 1&53 & 1&86 & 2&20 & 2&56 & 27&4 & 70&7 & \multicolumn{2}{c} {307} &
1&50 & 0&0810 \\
$^{230}$Th & 1&36(8)         & \omit\span      & \omit\span &
\omit\span & 5&7(26) & \omit\span & \multicolumn{2}{c} {20(11)} &
1&8(8) & 0&12(8) \\
           & 1&48 & 1&72 & 1&94 & 2&16  & 24&3 & 50&2 & \multicolumn{2}{c} {172} &
1&48 & 0&0797 \\
$^{232}$Th & 1&44(15) & 1&65(14) & 1&73(12) & 1&82(15) &
\multicolumn{2}{c}{14(6)} & 2&6(13) & \multicolumn{2}{c} {17(8)} &
2&48(42) & 0&045(20) \\
           & 1&46 & 1&68 & 1&85 & 2&03 & 21&9 & 42&0 & \multicolumn{2}{c}{130} &
1&46 & 0&0770 \\
$^{234}$U  & \omit\span &   \omit\span     & \omit\span
&\omit\span &\omit\span &\omit\span & \omit\span &
1&69(40) & 0&097(24) \\
           & 1&45 & 1&64 & 1&78 & 1&91 & 19&0 & 34&0 & \multicolumn{2}{c}{95} &
1&45 & 0&0751 \\
$^{236}$U  & 1&42(11) & 1&55(11) & 1&59(17) & 1&46(17) &
\omit\span & \omit\span & \omit\span & \omit\span
      &   \omit\span      \\
           & 1&45 & 1&63 & 1&76 & 1&88 & 17&8 & 31&3 & \multicolumn{2}{c}{85} &
1&45 & 0&0743 \\
$^{238}$U  & \omit\span       & \omit\span       & 1&45(23) &
1&71(22) & 1&4(6) & 3&6(14) & \multicolumn{2}{c}{12(5)} &
1&74(17) & 0&108(12) \\
           & 1&45 & 1&63 & 1&75 & 1&87 & 17&3 & 30&1 & \multicolumn{2}{c}{95} {80} &
1&45 & 0&0743 \\
$^{238}$Pu &  \omit\span            &  \omit\span     & \omit\span
& \omit\span& \multicolumn{2}{c}{14(4)} &\omit\span &
\multicolumn{2}{c}{11(4)} &
   \omit\span   &  \omit\span       \\
           & 1&45 & 1&63 & 1&75 & 1&86 & 17&2 & 29&8 &  \multicolumn{2}{c}{79} &
1&45 & 0&0741 \\
$^{250}$Cf &  \omit\span            &  \omit\span     & \omit\span
&\omit\span & \omit\span&\omit\span &\omit\span &
1&61(27) & 0&092(16) \\
           & 1&44 & 1&61 & 1&72 & 1&81 & 14&8 & 24&7 & \multicolumn{2}{c}{61} &
1&44 & 0&0730 \\
\hline
\end{tabular}
\end{table}

\begin{table}

\caption{Values of $\beta_0$ (determined through minimization of
$\sigma_{\beta,\gamma}$ [Eq. (\ref{eq:e32}]) and $C$ [determined
through Eq. (\ref{eq:e31})] corresponding to minimum rms deviation
$\sigma_{\beta,\gamma}$ [Eq. (\ref{eq:e32})] between the $\beta_1$
and $\gamma_1$ bands of the ES-D solution for fixed value of
$R_{0/2}$, when the even levels of both bands up to $L_{max}$ are
taken into account. See subsection 3.7 for further discussion.}

\bigskip

\begin{tabular}{ r r r r r }
\hline
$R_{0/2}$ & $L_{max}$ & $\beta_0$ & $C$ & $\sigma_{\beta,\gamma}$ \\
\hline
5.  & 10 & 0.00 &  1.0 & 5.026 \\
10. & 10 & 0.00 &  7.3 & 3.072 \\
15. & 10 & 1.94 & 12.9 & 1.141 \\
20. & 10 & 2.64 & 16.1 & 0.993 \\
25. & 10 & 3.12 & 19.4 & 0.907 \\
30. & 10 & 3.53 & 22.2 & 0.855 \\
    &    &      &      &       \\
5.  & 20 & 0.00 &  1.0 & 4.890 \\
10. & 20 & 0.00 &  7.3 & 4.528 \\
15. & 20 & 1.73 & 14.7 & 2.611 \\
20. & 20 & 2.57 & 17.7 & 2.531 \\
25. & 20 & 3.09 & 20.6 & 2.371 \\
30. & 20 & 3.50 & 23.9 & 2.197 \\
\hline
\end{tabular}
\end{table}

\begin{table}

\caption{Theoretical predictions of the ES-D solution for
$\beta_0=1.95$ and $C=12.9$ compared to experimental data for
$^{232}$Th \cite{NDS}. See subsection 3.7 for further discussion.}

\begin{tabular}{ r r r r r | r r r }
\hline
$L$& gsb & gsb & $\beta_1$ & $\beta_1$ & $L$ & $\gamma_1$ & $\gamma_1$ \\
   &    exp &  th    & exp    &  th    &    &    exp &  th    \\
\hline
 0 &  0.000 &  0.000 & 14.794 & 15.021 &  2 & 15.907 & 17.210 \\
 2 &  1.000 &  1.000 & 15.680 & 16.021 &  3 & 16.804 & 17.978 \\
 4 &  3.284 &  3.268 & 17.683 & 18.288 &  4 & 18.030 & 18.989 \\
 6 &  6.749 &  6.665 & 20.724 & 21.686 &  5 & 19.454 & 20.234 \\
 8 & 11.280 & 11.021 & 24.754 & 26.042 &  6 & 21.266 & 21.701 \\
10 & 16.751 & 16.161 & 29.762 & 31.182 &  7 & 23.213 & 23.379 \\
12 & 23.033 & 21.931 & 35.549 & 36.952 &  8 & 25.496 & 25.254 \\
14 & 30.035 & 28.200 & 42.138 & 43.221 &  9 & 27.750 & 27.313 \\
16 &        &        & 49.438 & 49.887 & 10 & 30.624 & 29.542 \\
18 &        &        & 57.356 & 56.868 & 11 & 33.219 & 31.929 \\
20 &        &        & 65.811 & 64.102 & 12 & 36.484 & 34.461 \\
\hline
\end{tabular}
\end{table}

\end{document}